\documentclass[twocolumn,preprintnumbers,showpacs,amsmath,amssymb,prb]{revtex4}

\providecommand{\e}[1]{\ensuremath{\times 10^{#1}}}
\usepackage{float}
\usepackage{graphicx}
\usepackage{dcolumn}
\usepackage{bm}
\usepackage{braket}
\usepackage{subfigure}

\newcommand{\Apar}{A_{\|}}
\newcommand{\Aperp}{A_{\bot}}

\newcommand{\lr}[1]{\left(  #1 \right)}

\newcommand{\abs}[1]{\left|  #1 \right|}

\makeatletter

\newcommand{\Rmnum}[1]{\expandafter\@slowromancap\romannumeral #1@}
\makeatother

\usepackage{hyperref}

\graphicspath{{Figures/}} 

\begin{document}

\title{Sidebands in Optically Detected Magnetic Resonance Signals of Nitrogen Vacancy Centers in Diamond}
\author{Maria Simanovskaia$^{1}$}
\author{Kasper Jensen$^{1}$}
\author{Andrey Jarmola$^{1}$}
\author{Kurt Aulenbacher$^2$}
\author{Neil Manson$^{3}$}
\author{Dmitry Budker$^{1,4}$}
\email{budker@berkeley.edu}
\affiliation{$^{1}$Department of Physics, University of California, Berkeley, CA 94720-7300, USA}
\affiliation{$^{2}$Institut f\"{u}r Kernphysik, Johannes Gutenberg-Universit\"{a}t Mainz, D-55099 Mainz, Germany}
\affiliation{$^{3}$Laser Physics Centre, Research School of Physics and Engineering, Australian National University, Australian Capital Territory 0200, Australia}
\affiliation{$^{4}$Nuclear Science Division, Lawrence Berkeley Laboratory, Berkeley, CA 94720, USA}

\begin{abstract}
We study features in the optically detected magnetic resonance (ODMR) signals associated with negatively charged nitrogen-vacancy (NV$^-$) centers coupled to other paramagnetic impurities in diamond. 
Our results are important for understanding ODMR line shapes and for optimization of devices based on NV$^-$ centers. We determine the origins of several side features to the unperturbed NV$^-$ magnetic resonance by studying their magnetic field and microwave power dependences. Side resonances separated by around 130 MHz are due to hyperfine coupling between NV$^-$ centers and nearest-neighbor $^{13}$C nuclear spins. Side resonances separated by approximately \{40, 260, 300\} MHz are found to originate from simultaneous spin flipping of NV$^-$ centers and single substitutional nitrogen atoms. All results are in agreement with the presented theoretical calculations. 
\end{abstract}


\maketitle

\section{Introduction}

	Negatively charged nitrogen-vacancy (NV$^-$) color centers in diamond are promising candidates for magnetometry with an unprecedented combination of sensitivity and spatial resolution~\cite{TAY2008}. In particular, diamonds with high density of NV$^-$ centers are appealing for ensemble magnetometry~\cite{ACO2009}. Optically detected magnetic resonance (ODMR) signals enable probing the energy levels of NV$^-$ centers, from which the external magnetic field can be inferred. In this work, we analyze lineshapes in ODMR signals of NV$^-$ centers in diamond. Specifically, we are able to separate features related to $^{13}$C hyperfine interactions from those related to simultaneous spin flips of NV$^-$ and single substitutional nitrogen atoms (P1 centers). At sufficiently high microwave powers, the weak coupling between the NV$^-$ and the P1 centers allows both spins to be flipped at the same time, absorbing one microwave photon.
	
	In the absence of magnetic fields, the NV$^-$ center has a magnetic resonance at a frequency of approximately 2870 MHz, which corresponds to a transition between the triplet ground-state magnetic sublevels with electron spin projections 
$m_S=0$ and  $m_S=\pm1$ 
[see Fig.~\ref{fig:NRG}]. 
Side resonances around this central resonance have been reported in the literature. These extra resonances are due to coupling between NV$^-$ centers and other paramagnetic impurities in the diamond lattice. 
	
	A pair of side resonances asymmetrically offset from the central peak and separated from each other by 130~MHz
has been attributed to hyperfine interaction between an NV$^-$ center and  a nearest-neighbor $^{13}$C nuclear spin~\cite{LOU1977,Bloch1985,FEL2009,MIZ2009,NIZ2010}. 
Hyperfine coupling with $^{13}$C spins located in other lattice sites has also been studied in detail~\cite{SME2011,Dreau2012}. 
Such interactions have been used to demonstrate quantum information processing using NV$^-$ centers and $^{13}$C nuclear spins \cite{JEL2004NUC,Neumann2008,Maurer2012}.
	
	Other side resonances have been observed in nitrogen-rich diamond~\cite{vanOort90,trial,JEN2012}. 
These resonances become particularly pronounced at high microwave powers, where the central resonance saturates, while these side resonances continue to grow with increasing microwave power. Understanding this regime is important for understanding light-narrowing effects in ODMR and for optimizing the performance of sensors based on ensembles of NV$^-$ centers in diamond~\cite{JEN2012}.
	
	At zero magnetic field, van Oort \emph{et al.}~\cite{vanOort90,trial} report pairs of side resonances separated by 140 and 280 MHz, symmetrically displaced from the central resonance.
The resonances separated by 140 MHz are unusually large compared to side resonances at 130 MHz (see Sec.~\ref{RandD}), and to our knowledge, they are reported only once in the literature. 
Van Oort \emph{et al.}~\cite{vanOort90} attribute these resonances to simultaneous spin flips of NV$^-$ centers and P1 centers.
Later, resonances separated by approximately 130 MHz, observed by Baranov \emph{et al.}~\cite{BAR2011} and Babunts \emph{et al.}~\cite{Babunts2012} in nanodiamonds, were interpreted as being related to coupling between NV$^-$ centers and substitutional nitrogen atoms. 
Baranov \emph{et al.} and Babunts \emph{et al.} assumed that these resonances are the same as those of van Oort \emph{et al.}; however, they are separated by 130 MHz, asymmetrically located around the central resonance, and are much smaller than those observed by van Oort \emph{et al.}, so they are more likely to be excellent examples of the side features associated with $^{13}$C.

	The P1 center has electron spin $1/2$ and nuclear spin 1. At zero field, these spins combine to form states with total angular momenta of $1/2$ and $3/2$ with a small splitting of the latter due to the absence of spherical symmetry in the diamond.
As a result, the level separations in the system are grossly unequal \{22, 127, and 149\} (as calculated in Sec.~\ref{P1theory}), in contrast to the equal separations incorrectly assumed by van Oort \emph{et al.}~\cite{vanOort90}

	In this paper we present improved ODMR spectra showing the different side resonances in diamonds with both high and low nitrogen concentration. 
The microwave-power and magnetic-field dependences of these features are studied in detail.
We are able to unambiguously ascribe the origins of all the observed side resonances by comparing experimental data with theoretical calculations.	
Furthermore, we show that the side resonances separated by 280 MHz observed in Ref.~\onlinecite{vanOort90} at zero magnetic field are partially unresolved pairs of resonances, one group separated by 260 MHz and another by 300 MHz.  
We determine the origin of side resonances separated by approximately 
\{40, 260, 300\} MHz to be simultaneous spin flips of the NV$^-$ centers and P1 centers, and we determine the origin of resonances separated by 130 MHz to be hyperfine coupling to $^{13}$C. 
We note that in measurements of all our samples, including the representative ones we present in this paper, we do not observe the side resonances split by 140 MHz reported by van Oort \emph{et al.}

\section{Methods}

\begin{figure}[t]
\begin{center}
	\subfigure[]{\includegraphics[width=42.5mm]{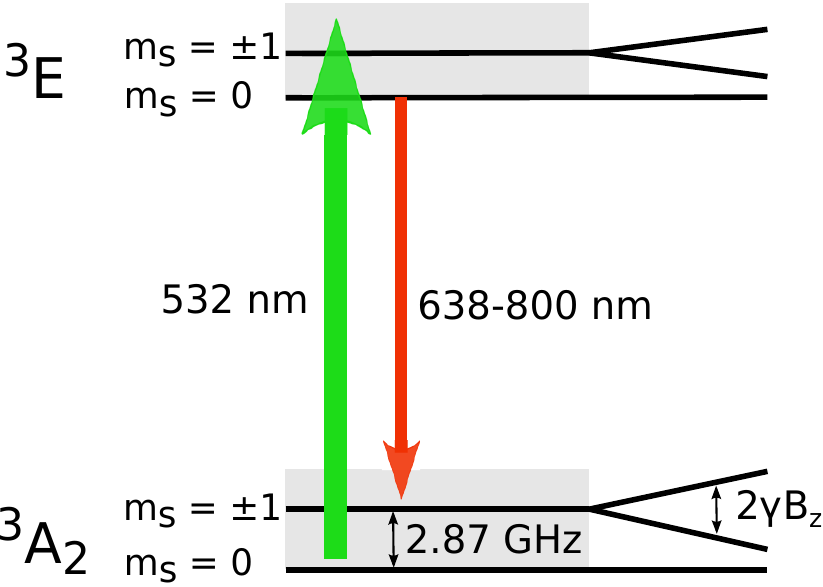}\label{fig:NRG}}
	\subfigure[]{\includegraphics[width=42.5mm]{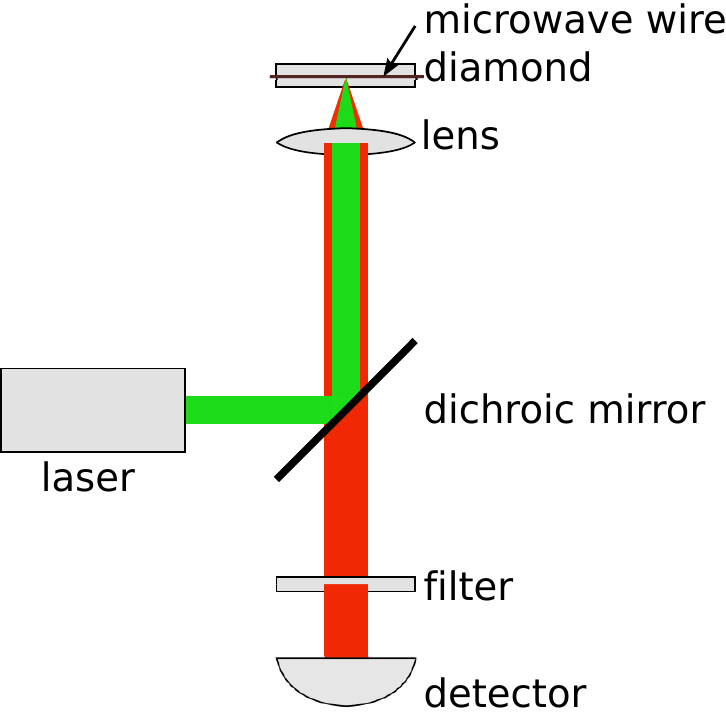}\label{fig:ODMR}}
\caption{(a) Energy-level structure of the NV$^-$ center. Gray areas represent phonon sidebands. $^3$A$_2$ and $^3$E are the spin-triplet ground and excited states, respectively. The electron spin projection along the NV$^-$ axis is denoted $m_S$. (b) Confocal-microscopy setup.} 
\end{center}
\end{figure}

	In our experiment, we used a conventional confocal-microscopy setup shown in Fig.~\ref{fig:ODMR}. We focus continuous-wave 532-nm laser light onto the diamond surface and optically pump the NV$^-$ centers to the $\ket{m_S = 0}$ ground-state sublevel. 
The fluorescence (wavelength $\approx$ 638-800 nm) propagates back through the lens, through a dichroic mirror, and onto a detector (photodiode or photomultiplier tube). Microwaves are applied with a wire pressed across the diamond surface. The microwave frequency is scanned over the transition between the $\ket{m_S = 0}$ and the $\ket{m_S = \pm1}$ sublevels. On resonance, the NV$^-$ centers are transferred to the $\ket{m_S = \pm 1}$ states, which leads to lower red fluorescence intensity than when the centers are in the $\ket{m_S = 0}$ state (see the discussion in Ref.~\onlinecite{ACO2010PRB}). We also measured Rabi oscillations by optically pumping NV$^-$ centers to the $\ket{m_S=0}$ state and applying resonant-microwave frequency lasting a few microseconds while keeping the laser on. This allows us to observe population oscillations between the $\ket{m_S = 0}$ and $\ket{m_S = \pm1}$ states in real time by monitoring fluorescence intensity modulation while microwaves are on. The microwave magnetic field that the NV$^-$ centers are exposed to depends not only on the microwave power but also on the distance from the transmitting wire. Microwave coupling strength is proportional to resonant microwave Rabi frequency, which is therefore used as a reference of coupling strength. We used low laser power in order to avoid optical pumping during the duration of oscillation.

	We tested diamond samples with different amounts of impurities. Single-crystal diamonds were obtained from Element Six with nitrogen concentrations of $\sim100$ ppm synthesized by the high-pressure, high-temperature technique (HPHT) or with nitrogen concentrations of $\sim1$ ppm synthesized by chemical-vapor deposition (CVD) \cite{BAL2009CVD}. 
Then, the samples were irradiated with relativistic electrons to create vacancies. In the measurements described here, we used the HPHT W5 and S8 samples
(S8 was previously studied in Ref.~\onlinecite{ACO2009}) and the CVD W2 sample, which were subjected to irradiation doses of $10^{18}$, $4\e{17}$, and $10^{16}$ cm$^{-2}$, respectively. W2 and W5 samples were irradiated at the Mainz Microtron facility in Germany. Finally, the diamond samples were annealed in vacuum for 3 h at $750^\circ$C to combine vacancies with nitrogen atoms to complete the formation of NV$^-$ centers. Based on measurements of the amount of fluorescence emitted from the two samples, we can estimate that the concentration of NV$^-$ centers is 25-100 times larger in the W5 sample compared to the W2 sample. The concentration of NV$^-$ centers in each sample varies by a factor of 2-4 depending on the exact spot on the diamond.

\section{Results and Discussion}
\label{RandD}

\begin{figure}[t]
\begin{center}
	\subfigure[]{\includegraphics[width=90mm]{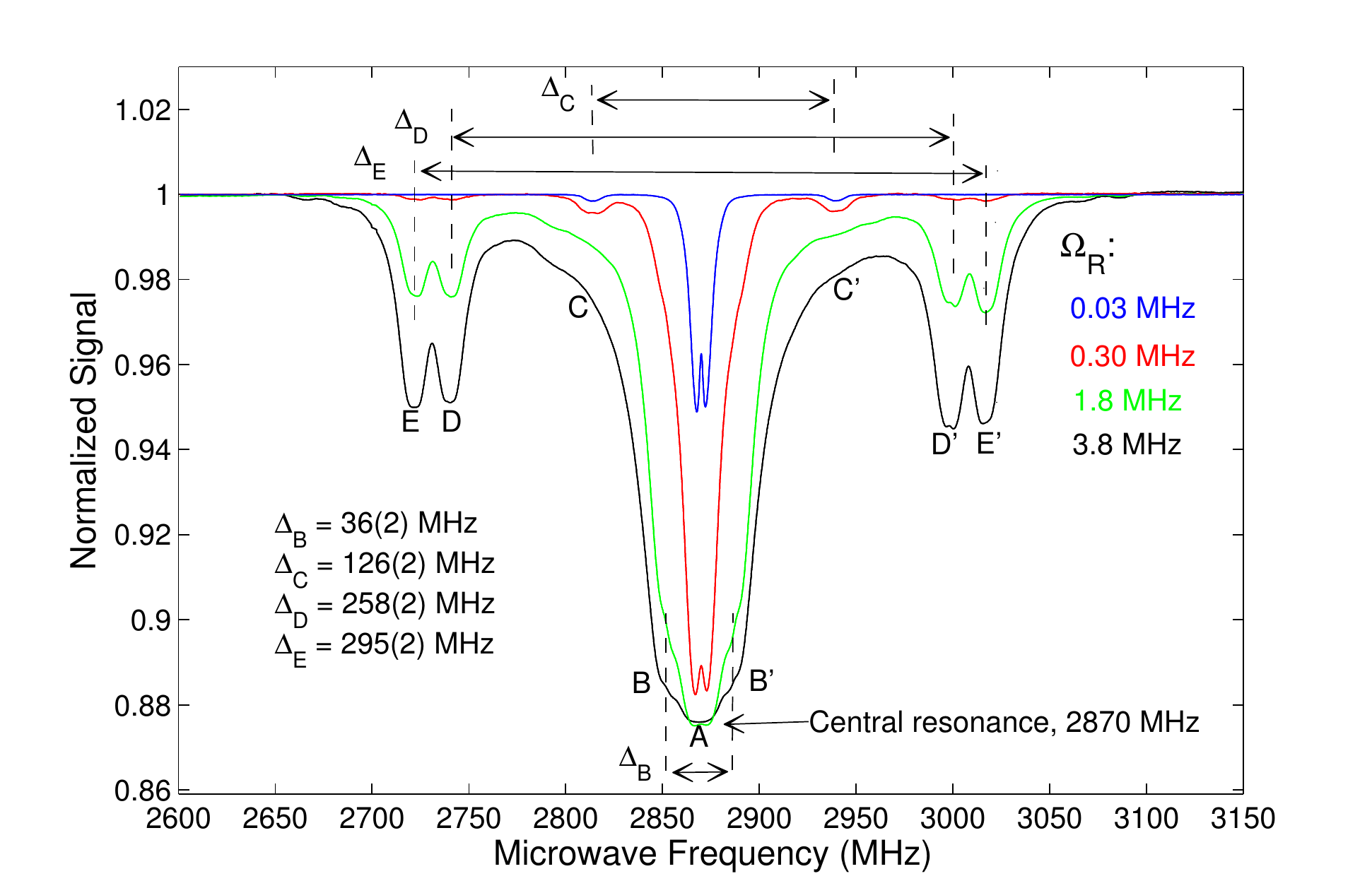}\label{fig:W5}}
	\subfigure[]{\includegraphics[width=90mm]{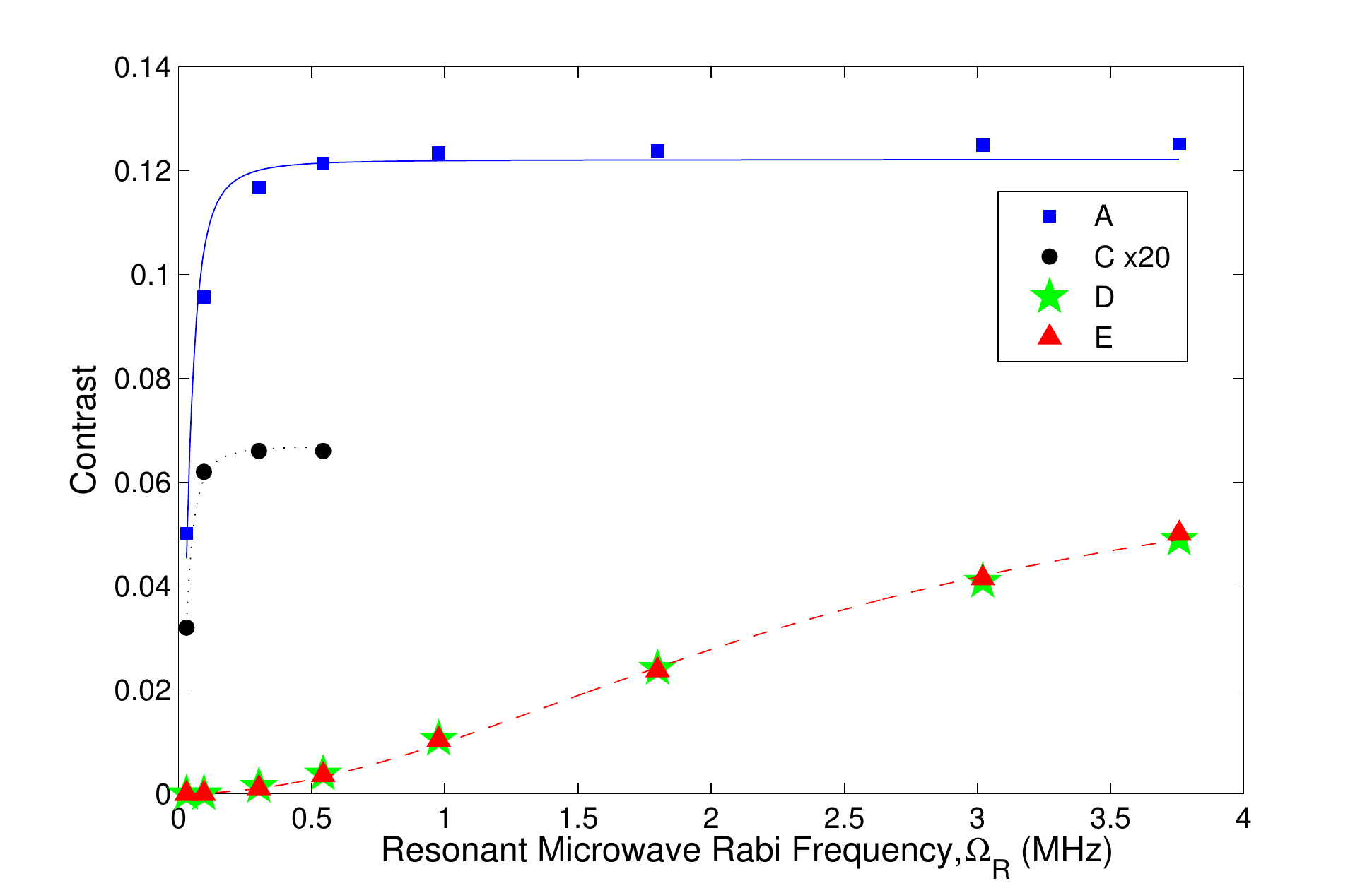}\label{fig:contrast}}
\caption{(a) Side-resonance features in ODMR are observed on either side of the central resonance at 2870 MHz. Measurements were done at zero magnetic field with a laser power of 0.7 mW, on sample W5. $\Omega_R$ denotes the resonant microwave Rabi frequency.
(b) Measurement contrast of side resonances A, C, D, and E is plotted as a function of $\Omega_R$, which is proportional to the square root of microwave power applied to the sample. The contrast of side-resonance group C is multiplied by 20 for clarity. We only plot the contrasts of the C resonances for resonant microwave Rabi frequencies $\Omega_R \leq0.6$ MHz, because the C resonances overlap with the B resonances for higher resonant microwave Rabi frequencies. The relative uncertainties of the contrasts are less than 5\%, which is smaller than the size of the markers on the plot.} 
\end{center}
\end{figure}

Figure~\ref{fig:W5} shows experimentally obtained ODMR spectra at zero magnetic field in the frequency range 2600 to 3150~MHz. At 2870~MHz, we observe the central resonance, which is labeled as resonance A in Fig.~\ref{fig:W5}. 
Resonance A corresponds to a transition between the $\ket{m_S=0}$ and $\ket{m_S=\pm1}$ states of the NV$^{-}$ center.
In measurements taken at small resonant microwave Rabi frequencies $\Omega_R$, which correspond to low microwave powers, the central resonance is split into two resonances because the presence of strain causes mixing between the $\ket{m_S=\pm1}$ sublevels and energy splitting between the resulting eigenstates [see, for example, the measurements taken at $\Omega_R=0.03$ MHz and $\Omega_R=0.30$ MHz in Fig.~\ref{fig:W5}]. Contrast, which is defined as the relative change of fluorescence intensity when the microwaves are on and off resonance with a transition, increases with increasing resonant microwave Rabi frequency. Figure~\ref{fig:contrast} shows the contrast of each resonance as a function of $\Omega_R$. 
The measured contrasts for each resonance were fitted to the function~\cite{JEN2012}
\begin{equation}
C(\Omega_R) = C_{sat} \, \frac{\left(\Omega_R/\Omega_{sat}\right)^2}{1+(\Omega_R/\Omega_{sat})^2},
\label{contrast}
\end{equation}
where $C(\Omega_R)$ is the contrast, $C_{sat}$ is the saturation contrast, and $\Omega_{sat}$ is the saturation Rabi frequency.
The measured contrasts and the fits are shown in Fig.~\ref{fig:contrast}, and the fitted parameters are shown in Table \ref{fitparameters}.
For resonance A, the saturation Rabi frequency was found to be 0.039(7) MHz.

\begin{figure}[t]
\begin{center}
	\subfigure[]{\includegraphics[width=50mm]{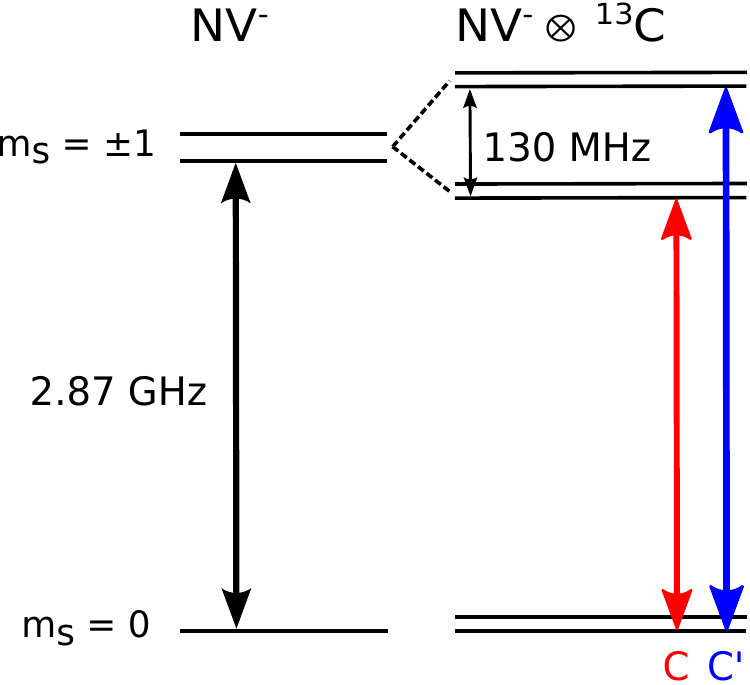}\label{fig:NV13C}}
	\subfigure[]{\includegraphics[width=75mm]{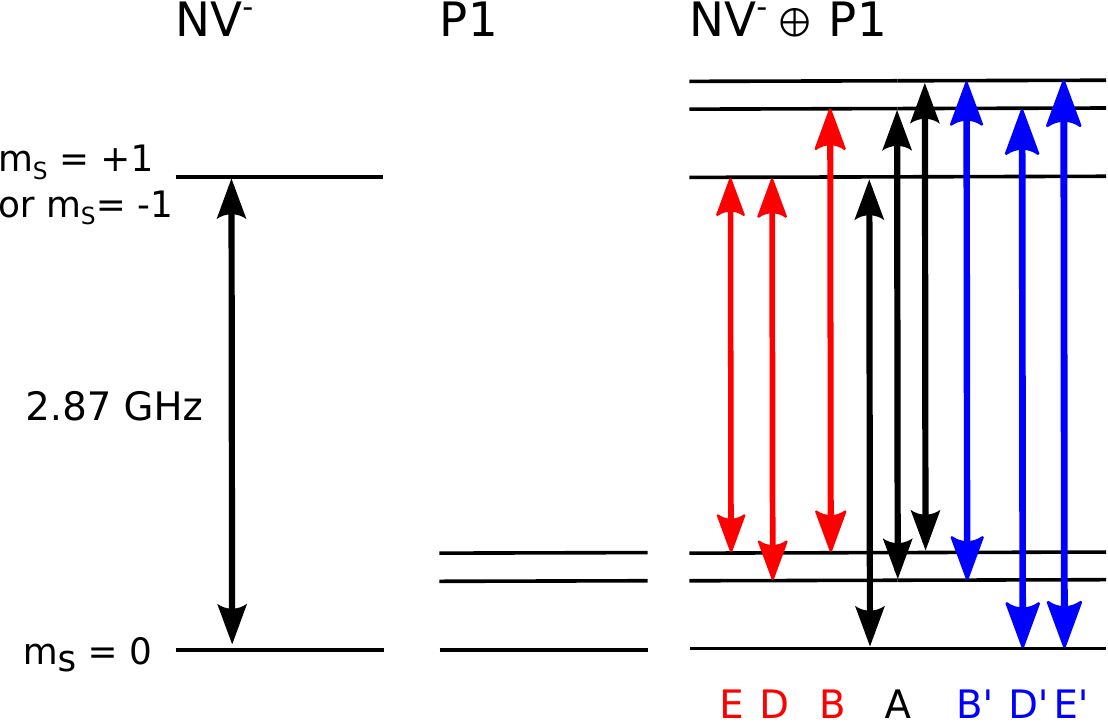}\label{fig:NVP1}}
\caption{Mixed energy-level diagrams for (a) hyperfine coupling of NV$^-$ centers to $^{13}$C nuclear spins and (b) simultaneous spin flipping of NV$^-$ centers and P1 centers. Each transition causing a resonance in Fig.~\ref{fig:W5} is labeled with the appropriate letter.}
\label{fig:levels}
\end{center}
\end{figure}

\begin{table}[t]
\caption{
Fit parameters obtained from a fit of the measured contrast as a function of resonant microwave Rabi frequency to Eq.~(\ref{contrast}). Data and fits are shown in Fig.~\ref{fig:contrast}.}
\begin{center}
\begin{tabular}{|c|c|c|} \hline
	Resonance & $C_{sat}$ & $\Omega_{sat}$ (MHz)\\ \hline
	Group A & 0.122(5) & 0.039(7) \\
	Group C & 0.0034(2) & 0.03(1) \\
	Group D & 0.072(5) & 2.4(2) \\
	Group E & 0.070(6) & 2.5(2) \\ \hline
\end{tabular}
\end{center}
\label{fitparameters}
\end{table}

	Apart from this central resonance, there are two small resonances in group C, $\Delta_C=126(2)\,\text{MHz}$, asymmetrically displaced to the low- and high- frequency sides of the center by around 56 and 70 MHz, respectively. For the lowest microwave power corresponding to $\Omega_R=0.03$ MHz, only side-resonance group C and the central resonance are observed [see Fig.~\ref{fig:W5}]. As seen in Fig.~\ref{fig:contrast} and from the fitted parameters in Table \ref{fitparameters}, the contrasts of resonance groups A and C saturate at similar values of the resonant microwave Rabi frequency. 
	This similar growth of contrasts of side-resonance groups A and C is expected, assuming that the C resonances are related to those NV$^-$ centers which have a nearest-neighbor $^{13}$C atom.
The NV$^-$ centers (electronic spin 1) and $^{13}$C atoms (nuclear spin 1/2) can interact through the hyperfine interaction, which is strongest when the $^{13}$C atom is in a nearest-neighbor position to the vacancy, where the electron density is the highest.
	The relative sizes of the contrasts can be estimated from the ratio of NV$^-$ centers with nearest-neighbor $^{13}$C atoms to those without nearest-neighbor $^{13}$C atoms. Since the natural abundance of $^{13}$C atoms is 1.1\% and there are three nearest-neighbor positions to the vacancy, we expect the ratio of contrasts to be roughly $(3 \times 0.011/2)=1.65\% \approx 1/60$. The factor of 2 appears because the ground state $\ket{m_S=0}$ splits into two states: $\ket{m_S=0,m_I=+1/2}$ and $\ket{m_S=0,m_I=-1/2}$ [as in Fig.~\ref{fig:NV13C}] due to the hyperfine interaction (here $m_S$ is the NV$^-$ electronic-spin projection and $m_I$ is the $^{13}$C nuclear-spin projection).
In the case that the $^{13}$C nuclear spins are unpolarized, the two ground states contain half of the population each. The measured ratio of the contrasts is found to be $3.0(3)\%$, which is a bit higher than the expected value because the side resonances are less affected by strain than  the central resonance.
	The reason these side resonances are less affected by strain can be understood from the fact that the magnetic field from the $^{13}$C nucleus splits the $m_S=1$ and $m_S=-1$ sublevels of the NV$^-$ center, making them insensitive, to first order, to the strain. A similar immunity to strain is observed when an external magnetic field is applied.

	At high microwave powers, one can also see resonances in group B with a separation of $\Delta_{B} = 36(2)\, \text{MHz}$, group D with a separation of $\Delta_{D}=258(2)\, \text{MHz}$, and group E with a separation of $\Delta_{E}=295(2)\, \text{MHz}$.
Side-resonance groups B, D, and E are symmetrically displaced from the center resonance.
The features in the W5 signal in Fig.~\ref{fig:W5} are too broad to resolve side-resonance group B; the value for $\Delta_B$ was therefore extracted from ODMR measurements of the S8 sample, which exhibits narrower features [Fig.~\ref{fig:4b}]. 
As seen in Fig.~\ref{fig:contrast} and Table \ref{fitparameters}, the contrast of resonance groups D and E requires higher microwave powers to saturate than resonance groups A and C. This implies that the D and E resonances are of different origins than the C resonances.\\

\subsection{Group C resonances}

\begin{figure}[t]
\begin{center}
	\subfigure[]{\includegraphics[width=90mm]{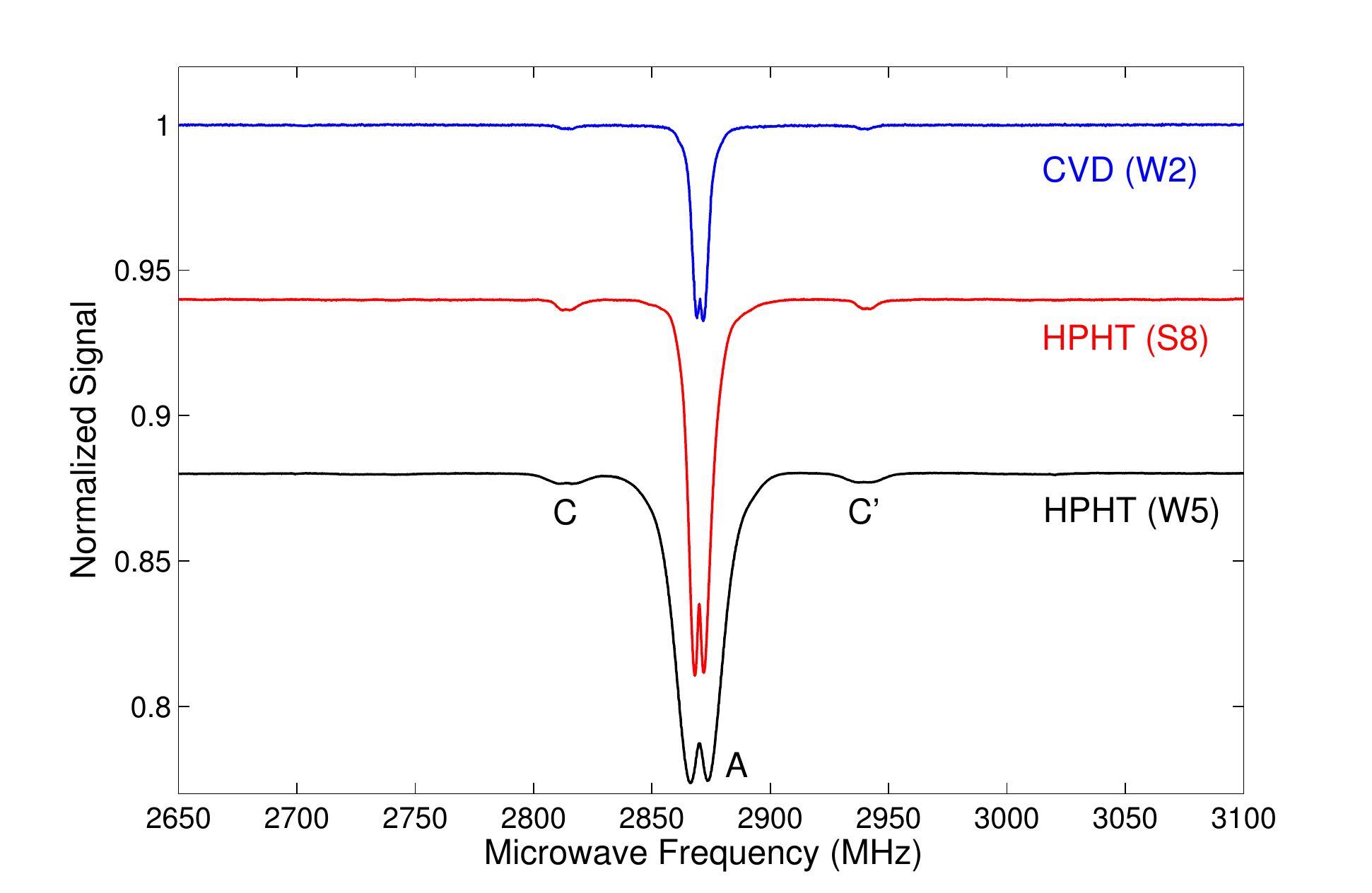}\label{fig:4a}}
	\subfigure[]{\includegraphics[width=90mm]{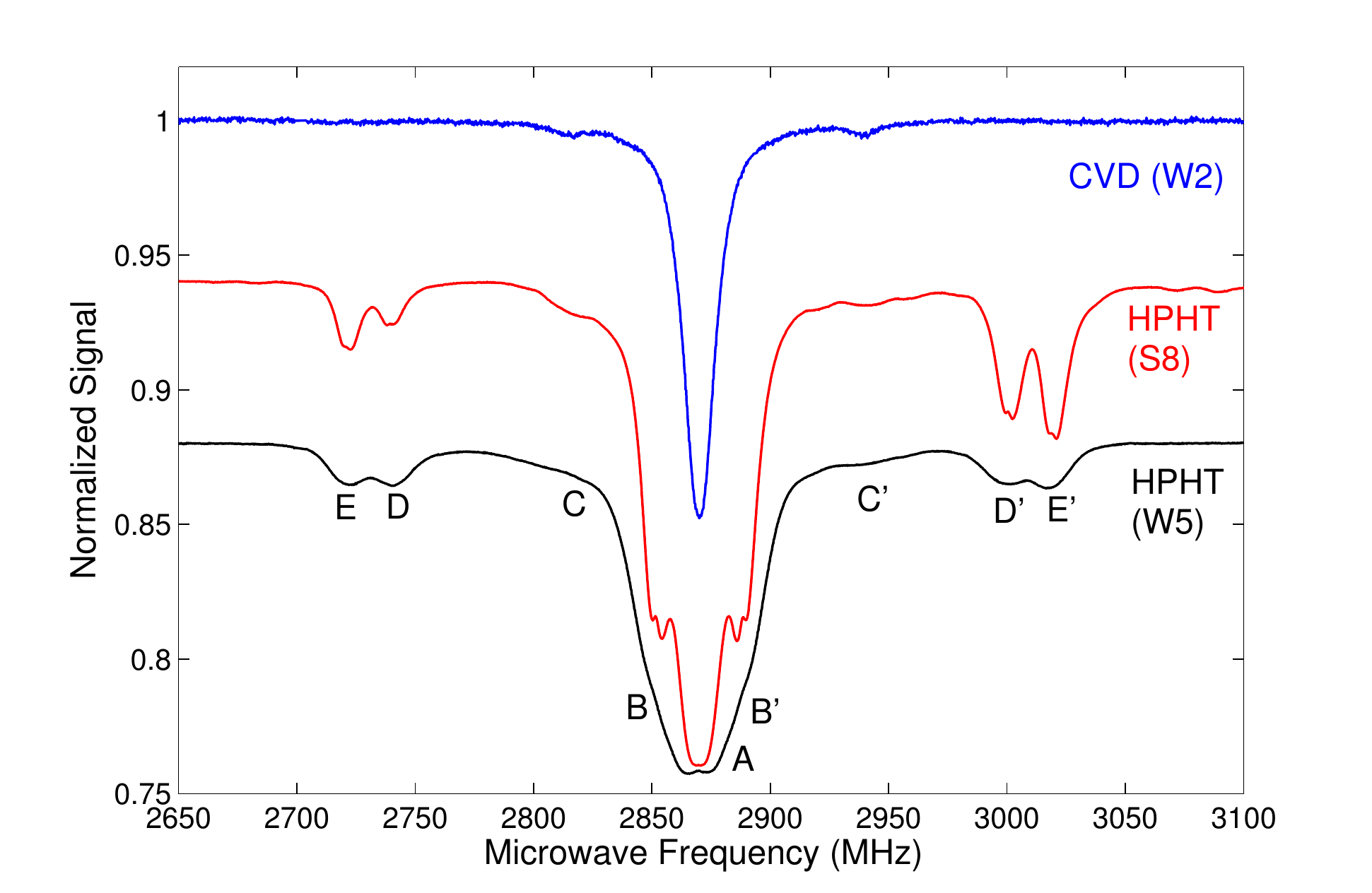}\label{fig:4b}}
\caption{Comparison of ODMR signals of CVD and HPHT diamonds. The nitrogen concentration is different between the samples, while the $^{13}$C concentration is the same. Measurements were done at zero magnetic field with a laser power of 1.7 mW,on diamond samples W2, and W5. The resonant microwave Rabi frequencies were (a) 0.35 and (b) 2.5 MHz. Measurements on the S8 sample were taken under slightly different experimental conditions than for the W2 and W5 samples. For the S8 measurements, the microwave power had some frequency dependence due to an etalon effect. }
\label{fig:4}
\end{center}
\end{figure}

	By measuring ODMR signals of samples with different concentrations of substitutional nitrogen and equal concentrations of $^{13}$C, we can gain information about the relevance of substitutional nitrogen and $^{13}$C to the signals. 
ODMR signals taken with the W2, S8, and W5 diamond samples are shown in Fig.~\ref{fig:4}. The essential differences between these three samples are the concentrations of substitutional nitrogen and NV$^{-}$ centers. The amount of $^{13}$C in the three samples is the same; it is set by the natural abundance of $^{13}$C of 1.1\%. Figure \ref{fig:4a} shows ODMR signals taken at low microwave power for the three samples. Side-resonance group C is present in all signals of the low-nitrogen-concentration sample W2 and those of the high-nitrogen-concentration samples W5 and S8. The ratios of contrasts between resonances in groups C and A are approximately 2.0\% and 3.0\% for the W2 and W5 samples, respectively. The difference in ratios is due to the varying amounts of strain in the samples. Figure \ref{fig:4b} shows ODMR signals taken at high microwave power. Side-resonance groups B, D, and E are present in the signals of the W5 and S8 samples and absent in that of the W2 sample. Since the concentration of $^{13}$C does not vary across samples and the side resonances in group C are present in the signals of all samples, the resonances in group C are most likely due to interactions of the NV$^-$ center with $^{13}$C. Moreover, this matches the previously observed side resonances with a separation of 130 MHz attributed to hyperfine coupling of the NV$^-$ center and a $^{13}$C nuclear spin~\cite{LOU1977,Bloch1985,FEL2009,MIZ2009}. The origin of the other side resonances possibly involves coupling of the NV$^-$ center to substitutional nitrogen, which is the highest-concentration impurity in the HPHT samples.\\
 
\noindent \emph{Hyperfine coupling to $^{13}$C nuclear spin}\\

	Besides presenting the experimental data, we also provide detailed theoretical calculations which predict the positions of all the side resonances. Here, we give a brief discussion of the calculations of relevance for side-resonance group C; for more details see Sec.~\ref{13Ctheory}. 

	As mentioned before, diamond consists of 1.1\% $^{13}$C, which has nuclear spin 1/2. When a $^{13}$C atom is located as a nearest neighbor to the vacancy, the hyperfine interaction between the NV$^-$ center electronic spin and the $^{13}$C nuclear spin is particularly strong. The hyperfine coupling can be written as
$\mathcal{H}_{\textrm{hf}} = \textbf{S} \cdot \bar{A} \cdot \textbf{I} $, where $\textbf{S}$ is the NV$^-$ center electronic spin, $\textbf{I}$ is the $^{13}$C nuclear spin, and $\bar{A}$ is the hyperfine tensor. 
As detailed in Sec.~\ref{13Ctheory} and illustrated in Fig.\ \ref{fig:NV13C}, we find that the energy eigenstates shift due to the hyperfine interaction. 
Using known values for the components of the hyperfine tensor, we calculate that magnetic resonance should occur at frequencies 
$(2870 - 56.9) = 2813.1$~MHz and $(2870 + 70.7) = 2940.7$~MHz. The separation between the side resonances is calculated to be 127.6(2)~MHz.
In an ensemble measurement, where only a fraction of the NV$^-$ centers has a nearest-neighbor $^{13}$C atom, we predict three magnetic resonances: one strong resonance with frequency around 2870 MHz and a pair of weaker side resonances separated by 127.6(2)~MHz. We finally note that the measured positions of the C side resonances agree with the calculated values to within a few megahertz, which supports that the C side resonances are related to the hyperfine interaction with $^{13}$C.\\

\noindent \emph{Magnetic-field dependence of group C resonances}\\

	In the presence of a magnetic field, the nominally degenerate $\ket{m_S=\pm1}$ states split according to the Zeeman effect. The splitting is illustrated in Fig.~\ref{fig:NRG} and is equal to $2\gamma B_z$, where $B_z$ is the projection of the external magnetic field on the NV$^-$ axis and $\gamma = 1.761\e{11}\, s^{-1}\, T^{-1}$ is the gyromagnetic ratio for the NV$^-$ center. Magnetic resonance between the $\ket{m_S=0}$ and $\ket{m_S=\pm1}$ states therefore occurs (to first order) at frequencies $D\pm \gamma B_z$, where $D \approx 2870$~MHz is the central resonance frequency at zero magnetic field. Note that there exist four possible orientations (along the four $[111]$ crystallographic directions) of the NV$^-$ center in the diamond lattice.  Furthermore, the magnitude of the projection $B_z$ on the NV$^-$ axis depends on the orientation of the NV$^-$ center. 

	We studied the magnetic-field dependence of all resonance group positions and observed different behaviors between groups C and \{B, D, E\}, which suggests different origins.
ODMR spectra of the W5 sample were taken for magnetic fields in the range from 0 to 200 G along the [100] crystallographic direction with high and low microwave powers (Fig.~\ref{fig:Bdep}). In this direction of applied magnetic field, all four orientations of the NV$^-$ center have equal projections $B_z$ of the magnetic field on the NV$^-$ axis. Figure~\ref{fig:Bchange2} shows low-microwave-power measurements in which we only see the group C resonances. At zero magnetic field, we observe the 130 MHz-split side resonances. When a magnetic field is applied along the [100] direction, these two side resonances split into a total of eight side resonances. 

	Figure~\ref{fit2} shows the difference between side resonance frequencies and the associated NV$^-$ central resonance as a function of magnetic field. We have also calculated the expected magnetic-field dependence of the magnetic resonances of an NV$^-$ center coupled to a nearest-neighbor $^{13}$C nuclear spin (see Sec.~\ref{13CBtheory} for details). Solid lines in Fig.~\ref{fit2} show the expected magnetic-field dependence. The theory describes the general trend of the resonance positions in a magnetic field. This confirms that the group C resonances are due to the interaction of the NV$^-$ center with $^{13}$C.

\subsection{Group B, D, and E resonances}

\begin{figure*}[t]
\begin{center}
	\subfigure[]{\includegraphics[width=80mm]{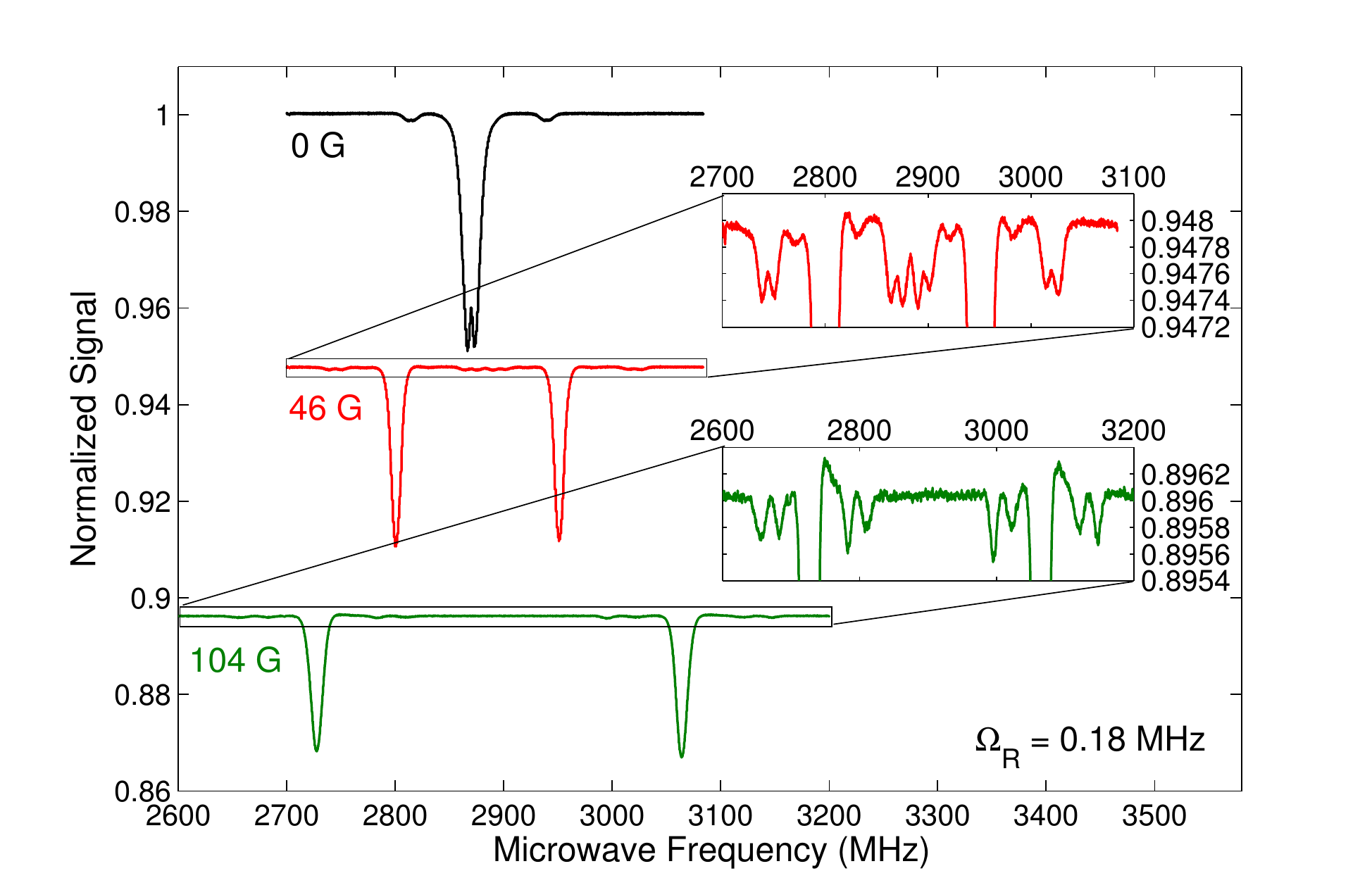}\label{fig:Bchange2}}
	\subfigure[]{\includegraphics[width=80mm]{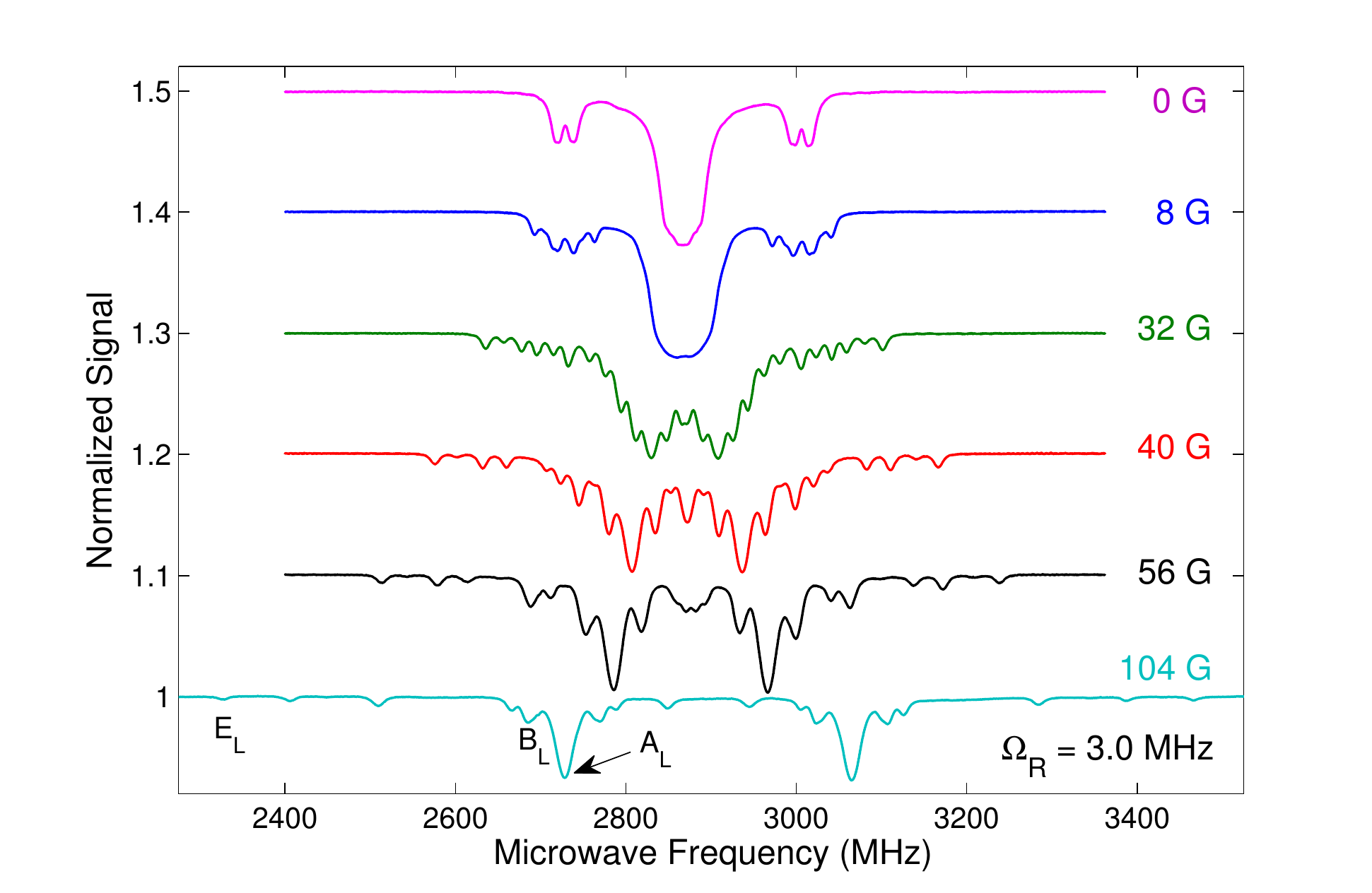}\label{fig:Bchange}}
	\subfigure[]{\includegraphics[width=80mm]{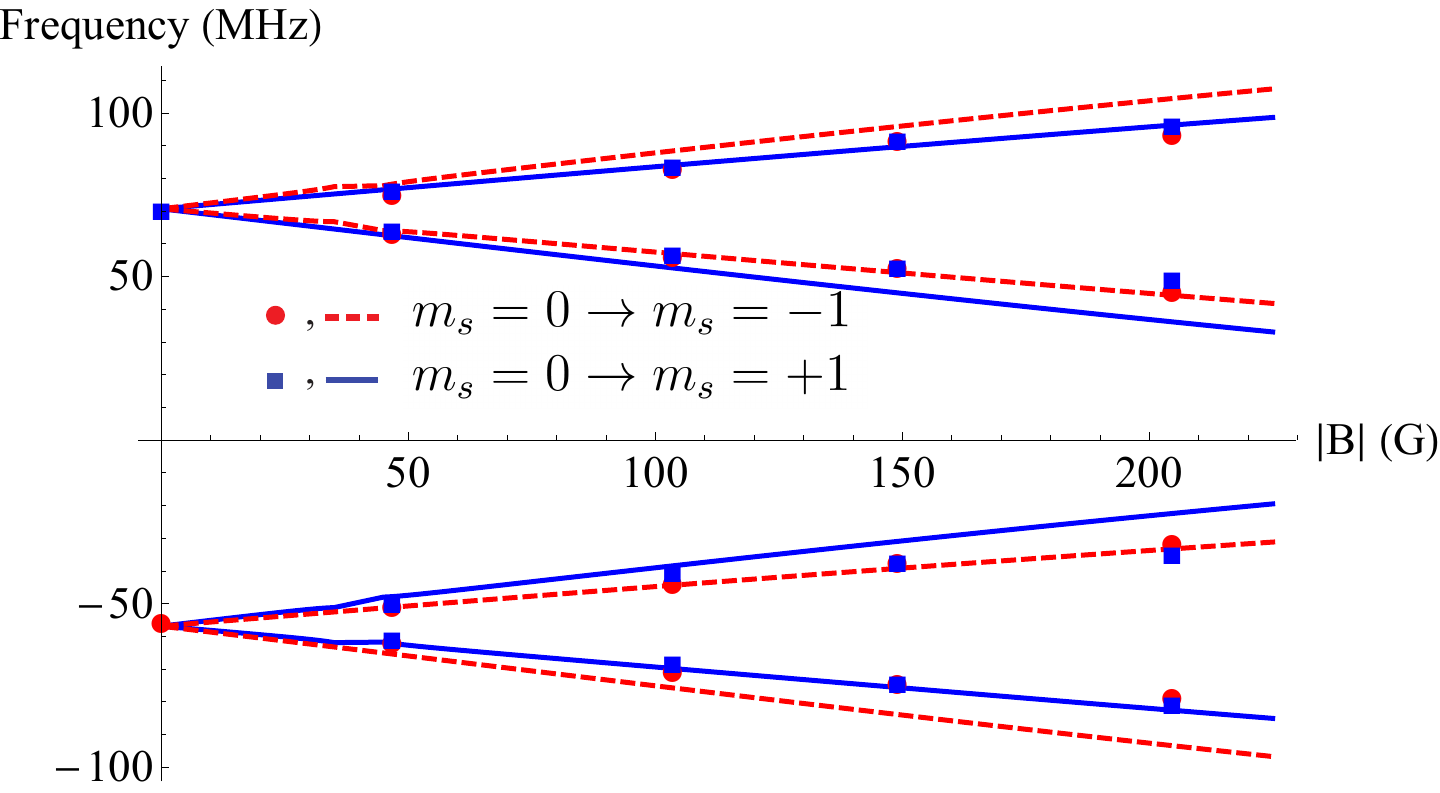}\label{fit2}}
	\subfigure[]{\includegraphics[width=80mm]{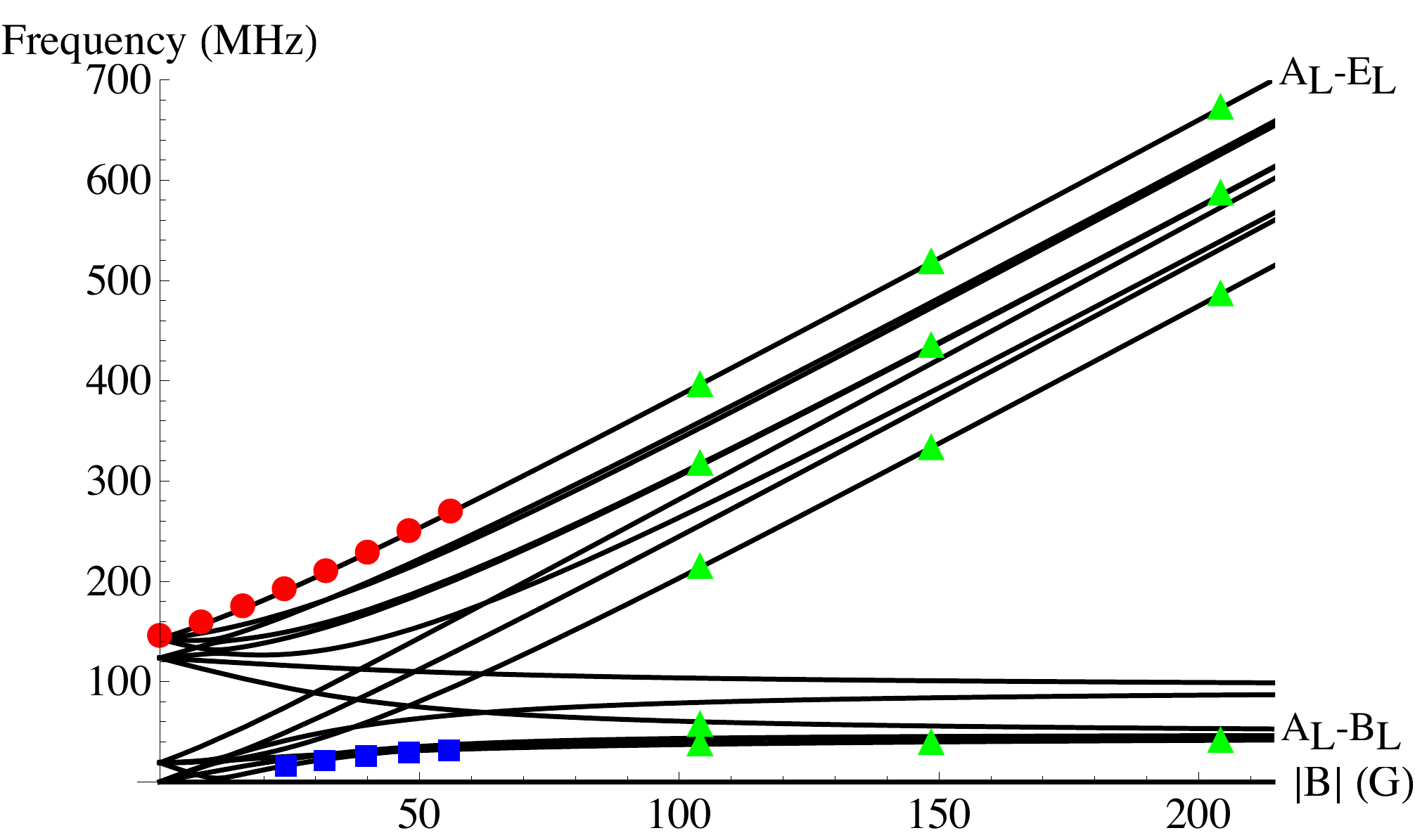}\label{fit}}
\caption{Magnetic-field dependence of side resonances for the W5 sample. A magnetic field was applied along the [100] orientation. ODMR measurements at low and high microwave powers corresponding to resonant microwave Rabi frequencies of (a) 0.18 and (b) 3.0 MHz. Signals are normalized and offset for clarity. Low-microwave-power measurements show C resonances, and high-microwave-power measurements show \{B, D, E\} resonances. The magnetic field strength is indicated near each curve. Note that the magnetic field along the NV$^-$ axis is $B_z=B\cos{54.7^\circ}.$ The values used for labeling purposes reflect the total $B$ value. (c) Frequency differences between group C resonances and their associated group A resonance as a function of applied magnetic field. Solid and dashed lines show theoretically predicted frequency differences for the $m_S=0 \rightarrow m_S=+1$ and $m_S=0 \rightarrow m_S=-1$ transitions, respectively. Points are experimental data from the measurements presented in (a). (d) Frequency differences between group \{B, D, E\} resonances and their associated group A resonance as a function of applied magnetic field. Solid lines show the theoretically predicted P1 center transition energies. Points are experimental data from the measurements presented in (b). Circles and squares represent frequency differences between group A resonances and their associated group E and group B side resonances, respectively. Triangles represent all resonance groups observed for magnetic fields higher than 100 G. The estimated uncertainties on the experimentally determined frequencies in (c) and (d) are 2 and 4~MHz, respectively.}
\label{fig:Bdep}
\end{center}
\end{figure*}
		
	We also studied spectra at high microwave power of the W5 diamond, which has a high concentration of nitrogen. As in Fig.~\ref{fig:Bchange}, in the presence of a magnetic field, we observe many more side resonances than at zero magnetic field. In spectra taken at magnetic fields lower than 60 G, we will focus on the side resonances farthest away (for example, $E_L$) and those closest ($B_L$) to the central resonance ($A_L$) for simplicity. The outermost side resonances belong to group E and the innermost side resonances belong to group B. At magnetic fields of 100-200 G, fewer side resonances are observed, so we can extract all of their positions. Their offsets from the associated center resonance $A$ are plotted  in Fig.~\ref{fit} together with their theoretically calculated positions in the presence of a [100]-oriented magnetic field.
The other lines in Fig.~\ref{fit} correspond to shifts of other resonances. 
The details of the calculations can be found in Secs.~\ref{P1theory}, \ref{P1NVtheory}, and \ref{Btheory}.
In the absence of an external magnetic field, the results of these calculations are $\Delta_B \approx 44\, \text{MHz}$, $\Delta_D \approx 254\, \text{MHz}$, and $\Delta_E \approx 298\, \text{MHz}$, which agree with the measured values to within a few megahertz. The model used for calculating these frequency separations has two input parameters which are related to the hyperfine interaction between the P1 center electronic spin ($S=1/2$) and the P1 center nuclear spin ($I=1$). The two parameters are the hyperfine interaction strength along the P1 center axis ($A_\parallel$) and perpendicular to the P1 center axis ($A_\perp$). In order to calculate the magnetic-field dependence, we also assume that the P1 center is oriented along one of the four [111] crystallographic directions.

	We have shown that the different side resonances can be distinguished by their dependence on microwave power and magnetic field strength. Side resonances due to interactions with P1 centers can only be observed at high microwave powers, while those due to hyperfine interaction with $^{13}$C are observed only at low microwave powers since they are masked by the other resonances in the W5 sample at high microwave powers. These measurements and the agreement of the positions as a function of the magnetic field of the side resonances with the theory derived in the Appendix further support interaction of the NV$^-$ center with the P1 center and the $^{13}$C nucleus as the origin for side-resonance groups \{B, D, E\} and C, respectively.\\

\begin{figure}[t]
\begin{center}
	\includegraphics[width=90mm]{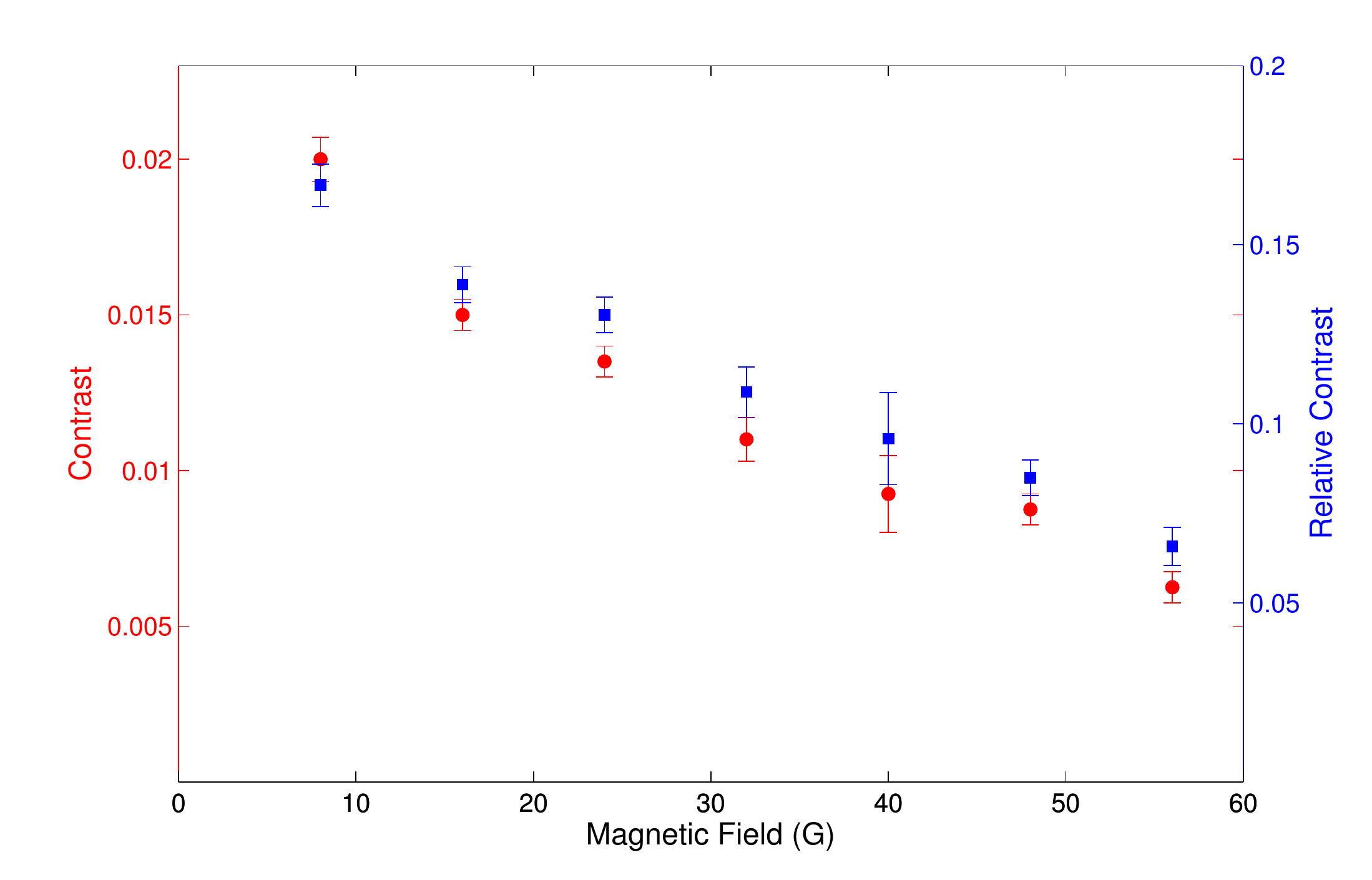}
\caption{Measurement contrast of side resonance E plotted as a function of magnetic field. Circles correspond to the contrast of the E resonances, and squares correspond to the contrast of the E resonances relative to that of the A resonances.  Data are aken at $\Omega_R = 3.0$~MHz.}
\label{fig:contrastVB}
\end{center}
\end{figure}

\newpage

\noindent \emph{Simultaneous spin flips of NV$^-$ and P1 centers}\\

	Apart from correctly predicting the positions of the resonances at various magnitudes of magnetic field, the theory discussed in Secs.~\ref{P1theory}, \ref{P1NVtheory}, and \ref{Btheory} gives us an idea of why the interaction between P1 and NV$^-$ centers leads to side resonances in the ODMR spectra. We provide a short discussion here. P1 centers have electronic spin 1/2 and nuclear spin 1. These impurities are present in HPHT-synthesized diamonds with concentration of $\sim100$ ppm, compared to the $^{13}$C concentration of $\sim10^4$ ppm. Due to the low concentration of P1 centers, the average separation between P1 and NV$^-$ centers is large, leading to a weak magnetic dipole-dipole coupling. This weak coupling slightly mixes the energy levels of the two, which allows simultaneous spin flips driven by the applied microwave field~\cite{vanOort90}. Such simultaneous spin flips result in side resonances displaced from the unperturbed NV$^-$ resonance by an amount corresponding to energies of transitions within the P1 center. In the absence of a magnetic field, the P1 center has three energy levels [each two fold degenerate; see Fig.~\ref{fig:NVP1}], leading to three transitions within the P1 center. The transition energies are calculated in detail in Secs.~\ref{P1theory}, \ref{P1NVtheory}, and \ref{Btheory} and are found to be $\{22,127,149\}$ MHz. This gives rise to three groups of sidebands symmetrically located around the unperturbed NV$^-$ resonance A with separations of $\{44, 254, 298\}$ MHz, as illustrated in Fig.~\ref{fig:NVP1}.
	
	As discussed previously, the dipole-dipole interactions mix the eigenstates of the combined NV/P1 system. The degree of mixing $x$ can be estimated as the ratio of the interaction strength to a typical energy separation of the states of the P1 center (which are the smaller energy intervals between the states that mix in the combined system).
The magnetic dipole-dipole interaction strength can be estimated to be around 1 MHz, assuming a density of 100 ppm of P1 centers (see Sec.~\ref{P1NVtheory}).
A typical energy separation of the states of the P1 center is around $100$ MHz  due to the hyperfine structure (see Fig.~\ref{Bdep}).
The mixing is then $x\approx 1 \ \mathrm{MHz} / \left( 100 \ \mathrm{MHz} \right) = 10^{-2}$.
When driving the combined system with microwaves, the transition amplitude for a side resonance should be approximately $x$ times the transition amplitude of the central resonance, i.e., approximately two orders of magnitude smaller.

In the experiment, the contrasts of the magnetic resonances are measured as a function of microwave power. The fitted saturation Rabi frequencies are expected to be proportional to the transition amplitudes of the particular transitions.
Experimentally, we find that the saturation Rabi frequencies of side resonances D and E are $\approx 60$ times larger than the saturation Rabi frequency of the central resonance (see $\Omega_{sat}$ in Table~\ref{fitparameters}). This is in agreement with our estimate.

Figure~\ref{fig:contrastVB} shows the contrast of the E resonances $C_E$ for magnetic fields in the range 0 to 56 G. The contrasts of the side resonances decrease with increasing magnetic field amplitude. Since the contrast of the A resonances $C_A$ decreases slightly with increasing magnetic field amplitude, we also plot their relative contrast ($C_E/C_A$). At approximately 50 G, the relative contrast has dropped by a factor of 2. At this magnetic field value, the Zeeman energy (2.8 MHz/G) becomes comparable to the hyperfine splittings of the P1 center, resulting in lower relative contrast since the mixing between the eigenstates of the combined NV/P1 system decreases.

\section{Conclusion and outlook}

	Experimentally and theoretically, we identify the origins of side resonances to the unperturbed NV$^-$ magnetic resonance which at zero magnetic field occurs at approximately 2870 MHz. We attribute symmetrically located side-resonance groups separated by around 40, 260, and 300 MHz to simultaneous flips of spins of the NV$^-$ center and weakly coupled P1 centers in the diamond. 
Our results confirm the assignment of the asymmetrically displaced side resonances separated by around 130 MHz to hyperfine interaction of the NV$^-$ center with a nearest-neighbor $^{13}$C nuclear spin. 
	
	The side resonances due to coupling between NV$^-$ and P1 centers are of importance for magnetic field sensing with NV$^-$ centers in nitrogen-rich diamonds. The existence of many side resonances complicates the ODMR spectrum and can lead to significant broadening of the NV$^-$ magnetic resonances when the side resonances overlap with the unperturbed NV$^-$ resonances. Simultaneous spin flips can be thought of as a source of relaxation which depolarizes the NV$^-$ center, leading to a reduced longitudinal spin relaxation ($T_1$) time. The reduction of $T_1$ was observed indirectly in Ref.~\onlinecite{JEN2012} as a nonlinear change in the linewidth of the NV$^-$ magnetic resonance as a function of resonant microwave Rabi frequency.
	
	The side resonances due to coupling of NV$^-$ and P1 centers could perhaps be used to polarize and detect P1 centers, which, in contrast to the NV$^-$ centers, cannot be directly optically polarized or probed. 
By driving several side resonances sequentially, it should be possible to fully polarize and initialize the P1 center in a specific quantum state. The ability to polarize P1 centers is important for magnetometry and quantum information processing with NV$^-$ centers. Previously, P1 centers have been thermally polarized by cooling a diamond located in a high magnetic field down to a few K~\cite{TAK2008}. Recently, a scheme of transferring spin polarization from NV$^-$ centers to nearby P1 centers by dressing their spin states with oscillating magnetic fields was demonstrated~\cite{BEL2013}.This polarization leads to improved spin-relaxation times $T_1$ and $T_2$, which are of importance for magnetic field sensing~\cite{JAR2012,PHA2012,STA2010}.

\section*{Acknowledgements}

	The authors of this paper are grateful to P. Kehayias, M. Ledbetter, B. Patton, V. M. Acosta, and D. English for useful discussions and help with the experiments. The authors also thank D. Suter and A. Gali for fruitful discussions. This work was supported by NSF Grant No. ECCS-1202258, the AFOSR/DARPA QuASAR program, IMOD, and the NATO SFP program. M.S. gratefully acknowledges support from a Summer Undergraduate Research Fellowship. K.J. acknowledges support from The Danish Council for Independent Research in Natural Sciences. D.B. acknowledges support by the Miller Institute for Basic Research in Science. 

\section{Appendix: Theory}

\subsection{Energy-level structure of the P1 center}	
\label{P1theory}	
	We here calculate the energy-level structure of the P1 center in diamond. 	The P1 center has electronic spin $S=1/2$ and nuclear spin $I=1$ which are coupled by the electron-nuclear hyperfine interaction $\mathcal{H}_\text{hf} = \mathbf{S}\cdot \bar{A} \cdot \mathbf{I}$ where $\bar{A}$ is the hyperfine-interaction tensor. Since the hyperfine interaction of a P1 center has axial symmetry along one of the four possible [111] crystallographic directions~\cite{SMI1959}, the hyperfine Hamiltonian can be written as
\begin{equation}
\mathcal{H}_\text{hf} = A_{\parallel}\cdot S_zI_z + A_{\perp}\cdot (S_xI_x + S_yI_y).
\end{equation} 
The $z$ direction is here defined as the direction where the P1 center hyperfine coupling is largest. We use the values $A_{\perp} = 81$ MHz and $A_{\parallel} = 114$ MHz, which have been measured using electron paramagnetic resonance~\cite{SMI1959} (uncertainties on $A_{\perp}$ and $A_{\parallel}$ are not given in Ref.~\onlinecite{SMI1959}).  
	
	First, we construct the ordered orthonormal tensor product basis $B$ in the form $\Ket{m_S, m_I}$ from the electron-spin basis $B_S=\{ \ket{-1/2},\ket{1/2} \}$ and the nuclear-spin basis $B_I=\{ \ket{-1},\ket{0},\ket{1} \}$, 
\begin{equation}
\small
 B =\left\{ \Ket{-\frac1{2},-1},\Ket{\frac1{2},-1},\Ket{-\frac1{2},0},\Ket{\frac1{2},0},\Ket{-\frac1{2},1},\Ket{\frac1{2},1} \right\}.  
\end{equation}
	Using this ordered basis, we find the following block-diagonal matrix:
\begin{equation}
\mathcal{H} =
\left( \begin{matrix}
	\frac{A_\parallel}{2} & 0 & 0 & 0 & 0 & 0\\
	0& -\frac{A_\parallel}{2} & \frac{A_\perp}{\sqrt{2}} & 0 & 0 & 0\\
	0 & \frac{A_\perp}{\sqrt{2}} & 0 & 0 & 0 & 0\\ 
	0 & 0 & 0 & 0 & \frac{A_\perp}{\sqrt{2}} & 0\\
	0 & 0 &  0 & \frac{A_\perp}{\sqrt{2}} & -\frac{A_\parallel}{2} & 0\\
	0 & 0 & 0 & 0 & 0 & \frac{A_\parallel}{2}\\
	\end{matrix} \right).
\end{equation}
The eigenenergies of this system are
\begin{equation}
\label{eigenenergiesP1}
\small
\left\{ \frac{A_\parallel}{2}, \frac{-A_\parallel+\sqrt{A_\parallel^2 + 8A_\perp^2}}{4}, \frac{-A_\parallel-\sqrt{A_\parallel^2 + 8A_\perp^2}}{4} \right\},
\end{equation}
each with a degeneracy of 2. Plugging in numerical values for $A_\parallel$ and $A_\perp$, we calculate the following values for the eigenenergies \{-92, 35, 57\} MHz. Between these states, there are three transitions with frequencies \{22, 127, 149\} MHz.

\subsection{Simultaneous spin flips of NV$^-$ and P1 centers}
\label{P1NVtheory}
	From experiment, we determined that resonance groups B, D, and E [see Fig.~\ref{fig:W5}] are due to coupling of the NV$^-$ center to P1 centers. More specifically, these resonances are due to simultaneous spin flips of the NV$^-$ center and a P1 center. The two impurities can interact through the magnetic dipole-dipole interaction. The  interaction strength depends on the distance between the NV$^-$ and P1 center and is expected to be weak (for instance, Ref.~\onlinecite{vanOort90} gives 1-10 MHz for the interaction strength). This dipole-dipole interaction allows for simultaneous spin-flips at frequencies corresponding to the sums (spin flip in the same direction) and differences (spin-flip in opposite directions) between the typical transitions of NV$^-$ centers (on the order of gigahertz) and the possible transitions of the P1 centers (on the order of zero to hundreds of megahertz) described above. 
	
	In the absence of strain and magnetic field, we predict resonances in the ODMR signal at frequencies of \{2721, 2743, 2848, 2870, 2892, 2997, 3019\} MHz [see Fig.~\ref{fig:NVP1}]. In other words, we predict a central resonance at 2870 MHz, and three sets of resonances symmetric about the center. These groups of side resonances should be separated by \{44, 254, 298\} MHz. We observe these resonances in Fig.~\ref{fig:W5} as groups B, D, and E, respectively. Furthermore, in Fig.~\ref{fig:4}, the resonances predicted to originate from the interaction of the NV$^-$ center with P1 centers are not observed in the ODMR signal of the CVD sample due to low concentration of P1 centers ($\sim1$ ppm). 
	
	We now estimate the magnetic dipole-dipole coupling strength between an NV center and a P1 center. The magnetic energy between two spins $\mathbf{S_1}$ and $\mathbf{S_2}$ can be written as
\begin{equation}
\small
\mathcal{H}_{\rm{dip}} = -\frac{\mu_0}{4 \pi} \frac{\gamma_1 \gamma_2}{r^3_{12}} \, \hbar^2 \, \left[ 3\left( \mathbf{S_1}\cdot \mathbf{\widehat{r}_{12}} \right)
\left( \mathbf{S_2}\cdot \mathbf{\widehat{r}_{12}} \right) - \mathbf{S_1}\cdot \mathbf{S_2}  \right] ,
\end{equation}
where $r_{12}$ is the distance between the two spins, $\mathbf{\widehat{r}_{12}}$ is a unit vector pointing from spin 1 to 2, $\gamma_1$ and $\gamma_2$ are the gyromagnetic ratios for spin 1 and 2, and $\mu_0$ is the vacuum permeability.
If we assume that the concentration of P1 centers is much higher than the concentration of NV centers, the average distance between an NV center and a P1 center is approximately equal to $n^{-1/3}$, where $n$ is the density of P1 centers. For a concentration of 100 ppm of P1 centers, we find $n = 1.76\e{19} \ \mathrm{cm}^{-3}$.
A characteristic magnitude of the dipole-dipole interaction energy is 
\begin{equation}
{E}_{\rm{dip}} \approx \frac{\mu_0}{4 \pi} \gamma_1 \gamma_2 n  \hbar^2 \times 1 \times 1/2,
\end{equation}
where 1 is the NV$^-$ electron spin and 1/2 is the P1 center electron spin.
Assuming gyromagnetic ratios $\gamma_1 = \gamma_2 = \gamma_e = 1.76\e{11} \ \mathrm{rad}/\left( \mathrm{s} \mathrm{T} \right)$ equal to the gyromagnetic ratio of the electron spin, we calculate
the dipole strength to be ${E}_{\rm{dip}}/h \approx$ 1 MHz.

\subsection{Magnetic-field dependence of energy-level structure of P1 centers}
\label{Btheory}
We now calculate the energy levels of the P1 center as a function of magnetic field. The zero-field energy levels were calculated above using the hyperfine Hamiltonian  $\mathcal{H}_{\rm{hf}}=\mathbf{S} \cdot \overline{A} \cdot \mathbf{I}$. The magnetic field can be described using the Hamiltonian 
$\mathcal{H}_B=\gamma \mathbf{B} \cdot \mathbf{S}$, where $\gamma$ is the gyromagnetic ratio. 
The total Hamiltonian is
\begin{eqnarray}
\mathcal{H}_{\rm{tot}}  &=& \mathcal{H}_{\rm{hf}}+\mathcal{H}_B \nonumber\\
	&= & \Apar \cdot S_z I_z + \Aperp \cdot \lr{ S_x I_x + S_y I_y}\nonumber\\
 	& +&\gamma \abs{B} \left[ \cos \lr{\phi} S_z + \sin \lr{\phi} S_x \right],
\end{eqnarray}
where $\phi$ is the angle between the direction of the magnetic field and the $z$ axis, which is defined as the direction along which the P1 center hyperfine interaction is largest, i.e., one of the [111] crystallographic directions. Notice that both the hyperfine interaction and the magnetic field give rise to preferred directions in space. The P1 center can be oriented along one of the four possible [111] directions. For a general magnetic field direction, the energy of the P1 center depends on its orientation. However, if the magnetic field is oriented along one of the four possible $[100]$ crystallographic directions, the P1 center energies are the same for all P1 centers, independent of their orientation. This is because the angle between any $[100]$ direction and any $[111]$ direction is $\phi \approx 54.74^\circ$.                                             

\begin{figure}[t]
\begin{center}
	\includegraphics[width=80mm]{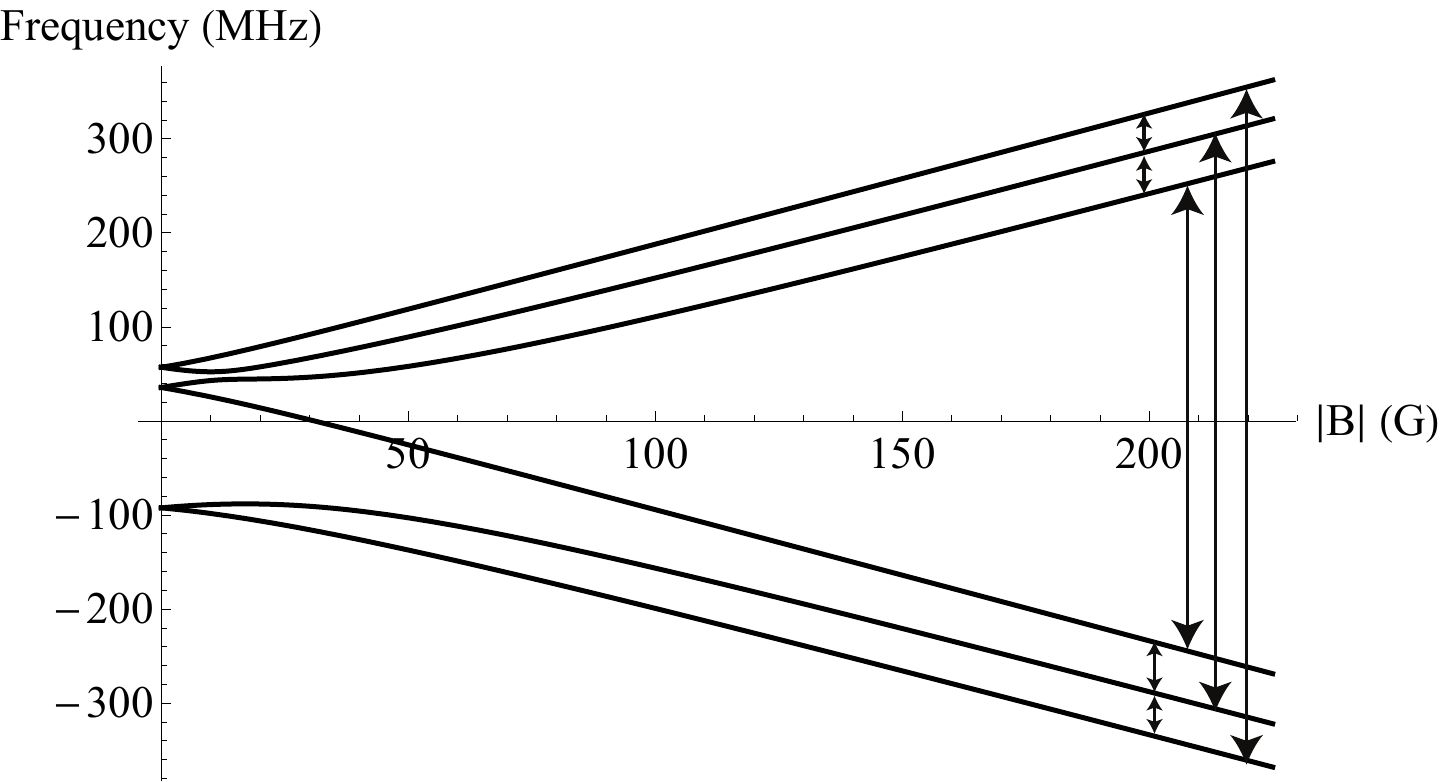}
\caption{The theoretically predicted dependence of the eigenenergy levels of the P1 center on the magnetic field applied in the [100] crystallographic direction.}
\label{Bdep}	
\end{center}
\end{figure}

	Diagonalizing the total Hamiltonian leads to the energy levels shown in Fig.~\ref{Bdep}.
At zero magnetic field, there are three energy levels which are each twofold degenerate. The zero-field energies are given in~(\ref{eigenenergiesP1}). When we consider nonzero magnetic fields, the degeneracies break to form six energy levels.
At high magnetic field, the energy levels split into two branches, corresponding to the P1 center electronic spin either parallel or antiparallel to the magnetic field.

	The possible transition frequencies can be calculated from the energy diagram shown in Fig.~\ref{Bdep}. With 6 energy levels, up to $15$ transitions are possible. The 15 possible transition frequencies are plotted in Fig.~\ref{fit} for magnetic field values up to 200 G (the range investigated in the experiment). Some transitions may be forbidden; for instance, when the magnetic field is large, we observe only four pairs of side resonances for each NV$^-$ central resonance. The transitions of the P1 center corresponding to these observed resonances are plotted as arrows in Fig.~\ref{Bdep}.
	
\subsection{Hyperfine coupling to $^{13}$C}
\label{13Ctheory}

	Diamond consists mainly of $^{12}$C atoms which have no nuclear spin. However, $^{13}$C, which is present due to the 1.1\% natural abundance, has nuclear
spin $I = 1/2$ and can couple to the NV$^-$ center electronic spin $S = 1$ through the magnetic dipolar (hyperfine) interaction. This interaction is particularly strong when the $^{13}$C atom is the nearest-neighbor to the vacancy in the NV$^-$ center. The nearest-neighbor hyperfine interaction between the NV$^-$ center and a $^{13}$C atom has been studied using electron paramagnetic resonance~\cite{LOU1977,FEL2009} and optically detected magnetic resonance~\cite{Bloch1985,MIZ2009}. 
	
	Here, we calculate the energy spectrum of the combined system consisting of the electronic spin of an NV$^-$ center and a nearest-neighbor $^{13}$C nuclear spin 
[see Fig.~\ref{fig:NV13C}]. In the absence of strain and magnetic fields, the NV$^-$ center has C$_{3v}$ symmetry. The ground states $\ket{m_S = 0}$ and $\ket{m_S = \pm  1}$ are separated by D $\approx$ 2.87 GHz. In this case, the Hamiltonian for the NV$^-$ center can be written as $\mathcal{H}_0 = D(S_z)^2$. The quantization axis for the NV$^-$ center, here denoted $z$, is along one of the four [111] crystallographic directions.
	
	The hyperfine interaction energy can be written as 
\begin{equation}
	\mathcal{H}_{\textrm{hf}} = \textbf{S} \cdot \bar{A} \cdot \textbf{I}  = \sum_{i,j={x,y,z}} S_i \bar{A}_{ij}I_{j}, 
\label{eq:13Chf}
\end{equation}
where $\bar{A}$ is the hyperfine tensor. For the case of the NV$^-$ center and a nearest-neighbor $^{13}$C atom, the hyperfine interaction has axial symmetry around the axis, here denoted by $z^\prime$, which connects the vacancy and the  $^{13}$C atom~\cite{LOU1977,GAL2008}. This axis is one of the [111] directions which is not the NV$^-$ axis. The angle between the $z$ and $z^\prime$ directions is $\theta \approx 109.47^\circ$.

	We consider the two coordinate systems $X$ and $X^\prime$ with axes $x, y, z$ and $x^\prime, y^\prime, z^\prime$. We choose the $y$ axes to coincide: $y = y^\prime$. $X^\prime$ can be obtained from $X$ by a rotation around the $y$ axis with angle $\theta$. We have the relation
\begin{equation}
\left( \begin{matrix} x^\prime \\ y^\prime \\ z^\prime \end{matrix} \right) = \left( \begin{matrix} \cos{\theta} & 0 & -\sin{\theta} \\ 0&1&0 \\ \sin{\theta} & 0 & \cos{\theta} \end{matrix} \right) \left(\begin{matrix}x\\y\\z\end{matrix}\right) \equiv \mathcal{R}(\theta)\left(\begin{matrix}x\\y\\z\end{matrix}\right).
\end{equation}	
We can express the hyperfine Hamiltonian in either coordinate system
\begin{equation}
\mathcal{H}_{\textrm{hf}} = \textbf{S}\cdot \bar{A}\cdot \textbf{I} = \textbf{S}^\prime \cdot \bar{A}^\prime \cdot \textbf{I}^\prime,
\end{equation}
where \textbf{S} and \textbf{I} are the electronic and nuclear spins, respectively, in the coordinate system $X$ and $\textbf{S}^\prime = \mathcal{R}(\theta)\textbf{S}$ and $\textbf{I}^\prime = \mathcal{R}(\theta)\textbf{I}$ in $X^\prime$.

	Since this particular hyperfine interaction is approximately axially symmetric, the hyperfine tensor has a simple form in the coordinate system $X^\prime$:
\begin{equation}
\bar{A}^\prime = \left(\begin{matrix}A_\perp&0&0\\0&A_\perp&0\\0&0&A_\parallel \end{matrix}\right).
\end{equation}
The hyperfine coupling parameters have been measured and calculated in previous works and have values $A_\parallel \approx$ 199.7(2) MHz and $A_\perp \approx$ 120.3(2) MHz ~\cite{FEL2009,GAL2008,LOU1977}.

	We can obtain the hyperfine tensor in the coordinate system $X$ using the transformation
\begin{equation}
\bar{A} = \mathcal{R}(\theta)^\dagger \bar{A}^\prime \mathcal{R}(\theta) = \left( \begin{matrix}A_{xx}&0&A_{xz} \\ 0&A_{yy}&0 \\ A_{zx}&0&A_{zz} \end{matrix} \right),
\end{equation}
where we have defined the constants $A_{xx} = A_\perp \cos^2{\theta}+A_\parallel \sin^2{\theta}$, $A_{yy}=A_\perp$, $A_{zz} = A_\parallel \cos^2{\theta}+A_\perp \sin^2{\theta}$ and $A_{xz} = A_{zx} = (A_\perp-A_\parallel)\sin{\theta}\cos{\theta}$. Using the above-mentioned values for $A_\parallel$, $A_\perp$, and $\theta$, we calculate $A_{xx} = 190.9(2)$~MHz, $A_{yy} = 120.3(2)$~MHz, $A_{zz} = 129.1(2)$~MHz and $A_{xz}=A_{zx}=-25.0(1)$~MHz. 

	We now find the eigenenergies of the combined system. Notice that the NV$^-$ zero-field splitting $D$ is much larger than the hyperfine-coupling parameters $A_\parallel$ and $A_\perp$. Using the secular approximation, where only the terms in $\mathcal{H}_{\textrm{hf}}$ which are proportional to $S_z$ are kept, we find the total Hamiltonian
\begin{equation}
 \mathcal{H}^{\textrm{sec}}_{\textrm{hf}} = \mathcal{H}_0+\mathcal{H}^{\textrm{sec}}_{\textrm{hf}} = D(S_z)^2+A_{zx} S_z I_x + A_{zz} S_z I_z.
\end{equation}
 The Hamiltonian can be written in matrix form using the basis $\{\ket{m_S,m_I}|m_S=-1,0,1,\,m_I=-1/2,+1/2\}$. By diagonalizing this Hamiltonian, we find the eigen-energies 0, $D\pm (A^2_{zx}+A^2_{zz} )^{1/2}/2$, each with a degeneracy of 2 [see Fig.~\ref{fig:NV13C}]. The combined system of an NV$^-$ center and a nearest-neighbor $^{13}$C atom has two magnetic resonances with frequencies $D\pm (A^2_{zx}+A^2_{zz})^{1/2}/2$. Inserting the values for $A_{zx}$ and $A_{zz}$, we find magnetic resonance frequencies of 2.87 GHz $\pm$ $(1/2) \times 131.5(2)$ MHz. Notice that with the secular approximation, one calculates that the two resonances should be symmetric around the unperturbed NV$^-$ resonance at $D \approx 2.87$ GHz.

	In the experiment, we observe that the resonance group C due to $^{13}$C is not symmetrically displaced from the central resonance A [see Fig.~\ref{fig:W5}]. This can be explained by corrections beyond the secular approximation.  We can diagonalize the full Hamiltonian $\mathcal{H}_{\mathrm{tot}} = \mathcal{H}_0 + \mathcal{H}_{\mathrm{hf}}$. In this case, we find the resonance frequencies to be $-56.9(1)$ and $+70.7(1)$~MHz displaced to the low- and high-frequency sides of the central resonance at around 2870~MHz. The separation between the two resonances is in this case 127.6(2)~MHz. Notice that by full diagonalization, we calculate that the resonances are asymmetrically located with respect to the unperturbed NV$^-$ resonance.

\begin{figure*}[t]
\begin{center}
	\subfigure[]{\includegraphics[width=80mm]{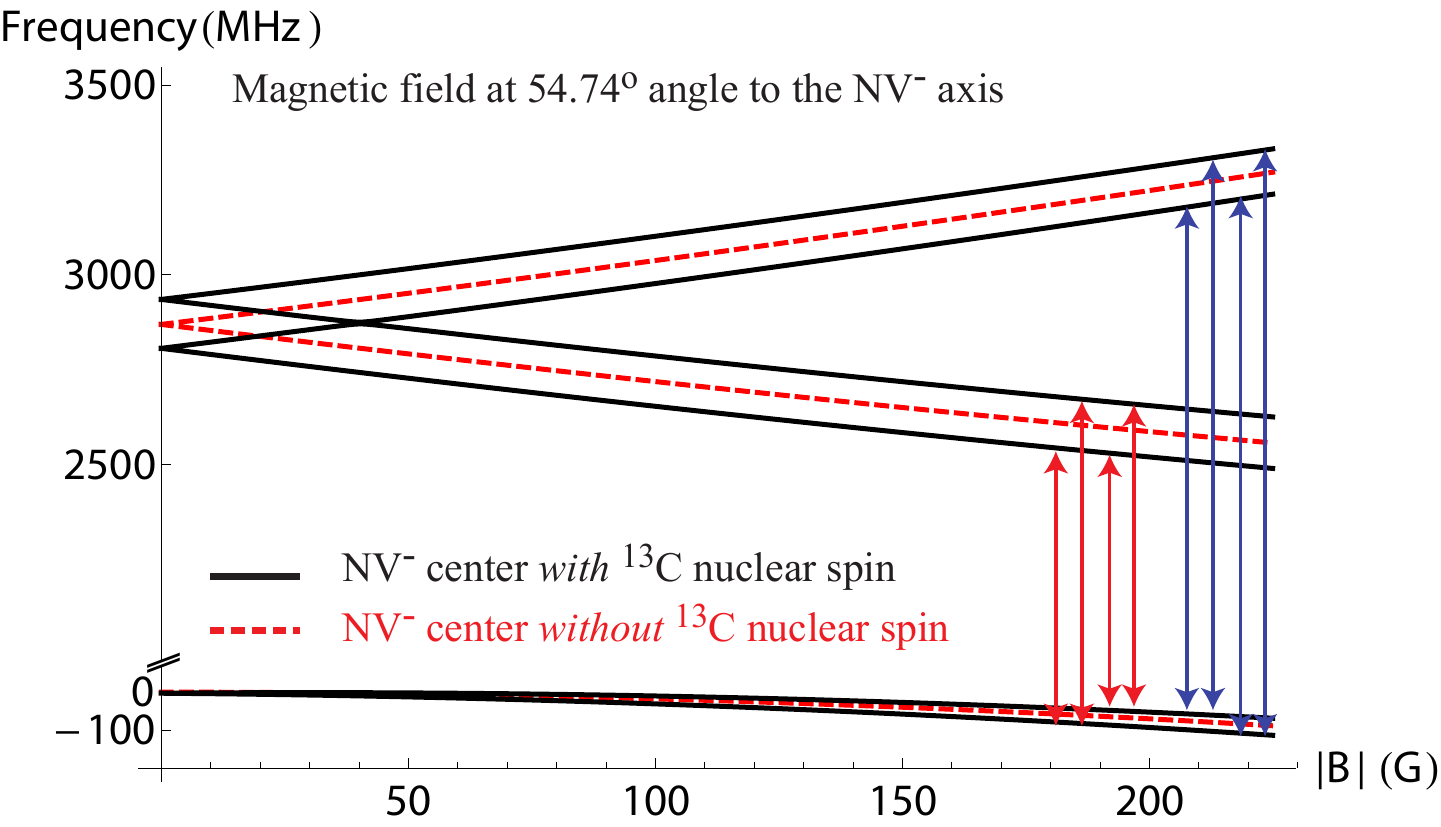}\label{BdepC13}}
	\subfigure[]{\includegraphics[width=80mm]{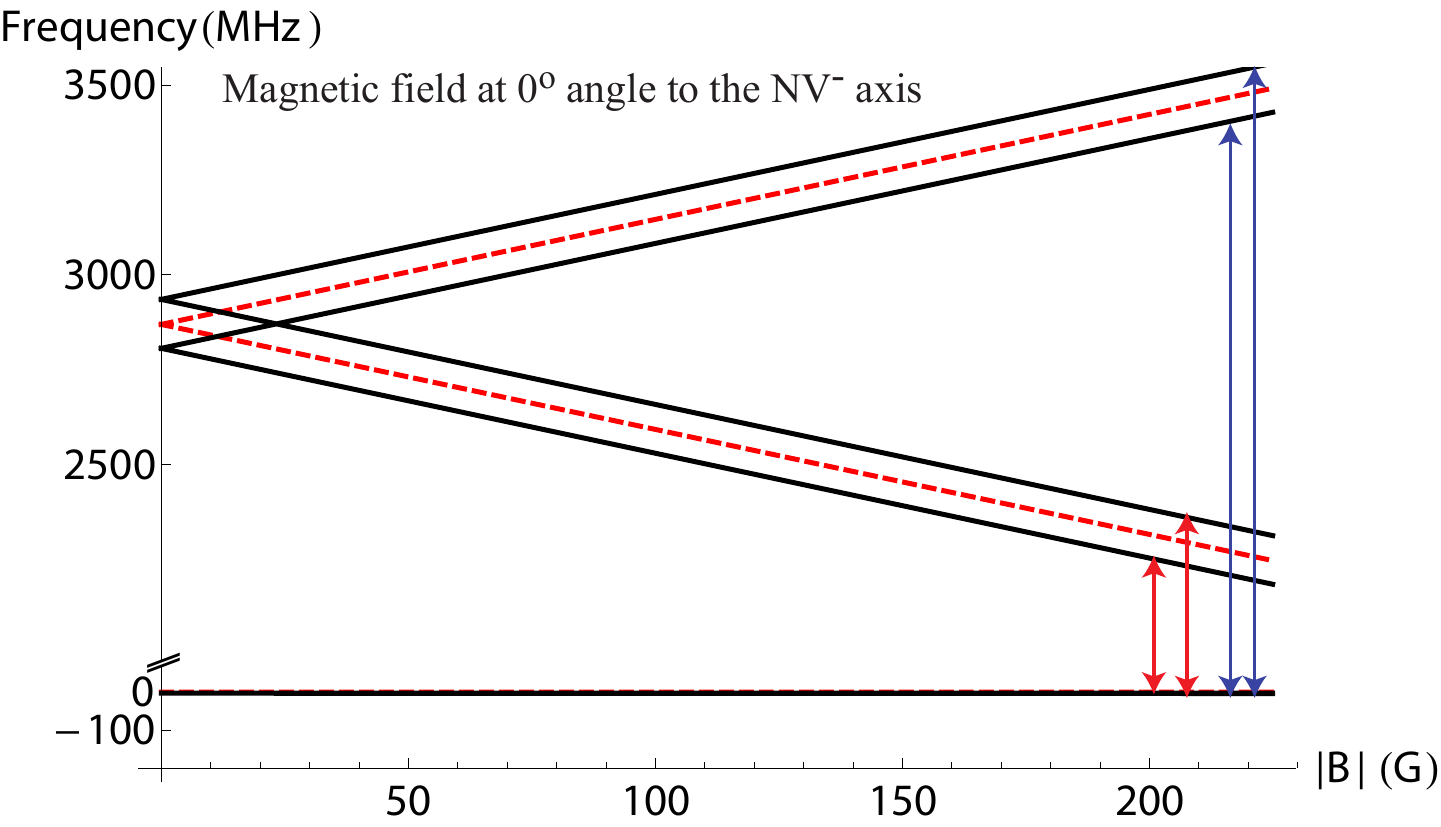}\label{BdepC13111}}
\caption{Dashed lines: energy level structure (three levels) of an NV$^-$ center located in a magnetic field applied along one of the (a) [100] crystallographic directions and (b) the NV$^-$ axis (which is one of the [111] directions) as a function of magnetic field strength. Solid lines: energy level structure of the combined system consisting of an NV$^-$ center and a nearest-neighbor $^{13}$C nuclear spin (six levels).}
\label{C13theory}	
\end{center}
\end{figure*}

\subsection{Magnetic-field dependence of side resonances due to $^{13}$C}
\label{13CBtheory}

We now calculate the energy-level structure of an NV$^-$ center and the combined system consisting of an NV$^-$ center and a nearest-neighbor $^{13}$C nuclear spin as a function of magnetic field strength.

The Hamiltonian describing an NV$^-$ center located in a magnetic field is 
\begin{equation}
\mathcal{H}_0 +\mathcal{H}_B = D \lr{S_z}^2+
\gamma \abs{B} \left[ \cos \lr{\phi} S_z + \sin \lr{\phi} S_x \right].
\end{equation}
If the magnetic field is applied along a [100] crystallographic direction, the angle between the NV$^-$ axis and the magnetic field is $\phi \approx 54.74^\circ$. The energy-level structure of the NV$^-$ center can be calculated by diagonalizing the Hamiltonian. The dashed lines in Fig.~\ref{BdepC13} show the calculated energies. The energy of the lowest level changes with magnetic field because the magnetic field is applied along the [100] direction, which differs from the direction of the NV$^-$ axis.

Some NV$^-$ centers have a nearest-neighbor $^{13}$C nuclear spin; in this case the Hamiltonian for the combined system is
$
\mathcal{H}_0 + \mathcal{H}_B + \mathcal{H}_{\rm{hf}},
$
where the hyperfine interaction energy is given by Eq.~(\ref{eq:13Chf}).
Note that we neglected the effect of the magnetic field on the $^{13}$C nuclear spin since the gyromagnetic ratio is much smaller for a nuclear spin than for an electronic spin. The energy level structure of the combined system calculated by diagonalizing the Hamiltonian is shown with solid lines in Fig.~\ref{BdepC13}.
We see that each of the energy levels of the NV$^-$ center splits into two due to the hyperfine interaction with the $^{13}$C nuclear spin.  
There is therefore a total of eight transitions between the two lower states and the four upper states [marked by arrows in Fig.~\ref{BdepC13}].
Measurements of ODMR spectra showing the C resonances are presented in Fig.~\ref{fig:Bchange2}. 
For magnetic fields in the range of 50 to 200 G, we observe eight side resonances.
Figure~\ref{fit2} shows the relative position of the C side resonances with respect to their associated A resonance (points: experiment, lines: theory). 
Based on the agreement between theory and experiment on the number of side resonances and on the resonance positions, we can conclude that the C side resonances originate from NV$^-$ centers \textit{with} a nearest-neighbor $^{13}$C nuclear spin. 

For completeness, we also present calculations of the NV$^-$ energy-level structure for the case where the magnetic field is aligned along the NV$^-$ axis.
In this case, the $m_S=0$ ground-state energy does not change with magnetic field [see Fig.~\ref{BdepC13111}, dashed lines].
Figure~\ref{BdepC13111} (solid lines) also shows the energy-level structure of NV$^{-}$ centers with a nearest-neighbor $^{13}$C nuclear spin. Note that the ground states $m_S=0,m_I=-1/2$, and $m_S=0,m_I=+1/2$ remain degenerate. In this case (when the magnetic field is aligned along the NV$^-$ axis), only four side resonances should be present. 


\begin{thebibliography}{22}
\expandafter\ifx\csname natexlab\endcsname\relax\def\natexlab#1{#1}\fi
\expandafter\ifx\csname bibnamefont\endcsname\relax
  \def\bibnamefont#1{#1}\fi
\expandafter\ifx\csname bibfnamefont\endcsname\relax
  \def\bibfnamefont#1{#1}\fi
\expandafter\ifx\csname citenamefont\endcsname\relax
  \def\citenamefont#1{#1}\fi
\expandafter\ifx\csname url\endcsname\relax
  \def\url#1{\texttt{#1}}\fi
\expandafter\ifx\csname urlprefix\endcsname\relax\def\urlprefix{URL }\fi
\providecommand{\bibinfo}[2]{#2}
\providecommand{\eprint}[2][]{\url{#2}}

\bibitem[{\citenamefont{Taylor et~al.}(2008)\citenamefont{Taylor, Cappellaro,
  Childress, Jiang, Budker, Hemmer, Yacoby, Walsworth, and Lukin}}]{TAY2008}
\bibinfo{author}{\bibfnamefont{J.~M.} \bibnamefont{Taylor}},
  \bibinfo{author}{\bibfnamefont{P.}~\bibnamefont{Cappellaro}},
  \bibinfo{author}{\bibfnamefont{L.}~\bibnamefont{Childress}},
  \bibinfo{author}{\bibfnamefont{L.}~\bibnamefont{Jiang}},
  \bibinfo{author}{\bibfnamefont{D.}~\bibnamefont{Budker}},
  \bibinfo{author}{\bibfnamefont{P.~R.} \bibnamefont{Hemmer}},
  \bibinfo{author}{\bibfnamefont{A.}~\bibnamefont{Yacoby}},
  \bibinfo{author}{\bibfnamefont{R.}~\bibnamefont{Walsworth}},
  \bibnamefont{and} \bibinfo{author}{\bibfnamefont{M.~D.} \bibnamefont{Lukin}},
  \bibinfo{journal}{Nat Phys} \textbf{\bibinfo{volume}{4}},
  \bibinfo{pages}{810} (\bibinfo{year}{2008}).

\bibitem[{\citenamefont{Acosta et~al.}(2009)\citenamefont{Acosta, Bauch,
  Ledbetter, Santori, Fu, Barclay, Beausoleil, Linget, Roch, Treussart
  et~al.}}]{ACO2009}
\bibinfo{author}{\bibfnamefont{V.~M.} \bibnamefont{Acosta}},
  \bibinfo{author}{\bibfnamefont{E.}~\bibnamefont{Bauch}},
  \bibinfo{author}{\bibfnamefont{M.~P.} \bibnamefont{Ledbetter}},
  \bibinfo{author}{\bibfnamefont{C.}~\bibnamefont{Santori}},
  \bibinfo{author}{\bibfnamefont{K.~M.~C.} \bibnamefont{Fu}},
  \bibinfo{author}{\bibfnamefont{P.~E.} \bibnamefont{Barclay}},
  \bibinfo{author}{\bibfnamefont{R.~G.} \bibnamefont{Beausoleil}},
  \bibinfo{author}{\bibfnamefont{H.}~\bibnamefont{Linget}},
  \bibinfo{author}{\bibfnamefont{J.~F.} \bibnamefont{Roch}},
  \bibinfo{author}{\bibfnamefont{F.}~\bibnamefont{Treussart}},
  \bibnamefont{et~al.}, \bibinfo{journal}{Physical Review B}
  \textbf{\bibinfo{volume}{80}}, \bibinfo{pages}{115202}
  (\bibinfo{year}{2009}).

\bibitem[{\citenamefont{Loubser and van Wyk}()}]{LOU1977}
\bibinfo{author}{\bibfnamefont{J.~H.~N.} \bibnamefont{Loubser}}
  \bibnamefont{and} \bibinfo{author}{\bibfnamefont{J.~A.} \bibnamefont{van
  Wyk}}, \bibinfo{note}{in \textit{Diamond Research} 1977, edited by P. Daniel,
  De Beers Industrial Diamond Division, Ascot, 1977, pp.11-14}.

\bibitem[{\citenamefont{Bloch et~al.}()\citenamefont{Bloch, Brocklesby, Harley,
  and Henderson}}]{Bloch1985}
\bibinfo{author}{\bibfnamefont{P.~D.} \bibnamefont{Bloch}},
  \bibinfo{author}{\bibfnamefont{W.~S.} \bibnamefont{Brocklesby}},
  \bibinfo{author}{\bibfnamefont{R.~T.} \bibnamefont{Harley}},
  \bibnamefont{and} \bibinfo{author}{\bibfnamefont{M.~J.}
  \bibnamefont{Henderson}}, \emph{\bibinfo{title}{Effects of microwave
  excitation on spectral hole-burning in colour centre systems}},
  \bibinfo{note}{{Journal de physique, Colloque C7, suppl\'{e}ment au
  n$^\circ$10, Tome 46, pp. 527-530, octobre 1985}}.

\bibitem[{\citenamefont{Felton et~al.}(2009)\citenamefont{Felton, Edmonds,
  Newton, Martineau, Fisher, Twitchen, and Baker}}]{FEL2009}
\bibinfo{author}{\bibfnamefont{S.}~\bibnamefont{Felton}},
  \bibinfo{author}{\bibfnamefont{A.~M.} \bibnamefont{Edmonds}},
  \bibinfo{author}{\bibfnamefont{M.~E.} \bibnamefont{Newton}},
  \bibinfo{author}{\bibfnamefont{P.~M.} \bibnamefont{Martineau}},
  \bibinfo{author}{\bibfnamefont{D.}~\bibnamefont{Fisher}},
  \bibinfo{author}{\bibfnamefont{D.~J.} \bibnamefont{Twitchen}},
  \bibnamefont{and} \bibinfo{author}{\bibfnamefont{J.~M.} \bibnamefont{Baker}},
  \bibinfo{journal}{Physical Review B} \textbf{\bibinfo{volume}{79}},
  \bibinfo{pages}{075203} (\bibinfo{year}{2009}).

\bibitem[{\citenamefont{Mizuochi et~al.}(2009)\citenamefont{Mizuochi, Neumann,
  Rempp, Beck, Jacques, Siyushev, Nakamura, Twitchen, Watanabe, Yamasaki
  et~al.}}]{MIZ2009}
\bibinfo{author}{\bibfnamefont{N.}~\bibnamefont{Mizuochi}},
  \bibinfo{author}{\bibfnamefont{P.}~\bibnamefont{Neumann}},
  \bibinfo{author}{\bibfnamefont{F.}~\bibnamefont{Rempp}},
  \bibinfo{author}{\bibfnamefont{J.}~\bibnamefont{Beck}},
  \bibinfo{author}{\bibfnamefont{V.}~\bibnamefont{Jacques}},
  \bibinfo{author}{\bibfnamefont{P.}~\bibnamefont{Siyushev}},
  \bibinfo{author}{\bibfnamefont{K.}~\bibnamefont{Nakamura}},
  \bibinfo{author}{\bibfnamefont{D.~J.} \bibnamefont{Twitchen}},
  \bibinfo{author}{\bibfnamefont{H.}~\bibnamefont{Watanabe}},
  \bibinfo{author}{\bibfnamefont{S.}~\bibnamefont{Yamasaki}},
  \bibnamefont{et~al.}, \bibinfo{journal}{Physical Review B}
  \textbf{\bibinfo{volume}{80}}, \bibinfo{pages}{041201}
  (\bibinfo{year}{2009}).

\bibitem[{\citenamefont{Nizovtsev et~al.}(2010)\citenamefont{Nizovtsev, Kilin, Neumann, Jelezko, and Wrachtrup}}]{NIZ2010}
\bibinfo{author}{\bibfnamefont{A.P.}~\bibnamefont{Nizovtsev}},
  \bibinfo{author}{\bibfnamefont{S.Ya.}~\bibnamefont{Kilin}},
  \bibinfo{author}{\bibfnamefont{P.}~\bibnamefont{Neumann}},
  \bibinfo{author}{\bibfnamefont{F.}~\bibnamefont{Jelezko}},
  \bibnamefont{and} \bibinfo{author}{\bibfnamefont{J.}~\bibnamefont{Wrachtrup}},
  \bibinfo{journal}{Optics and Spectroscopy} \textbf{\bibinfo{volume}{108}},
  \bibinfo{pages}{239} (\bibinfo{year}{2010}).

\bibitem[{\citenamefont{Smeltzer et~al.}(2011)\citenamefont{Smeltzer,
  Childress, and Gali}}]{SME2011}
\bibinfo{author}{\bibfnamefont{B.}~\bibnamefont{Smeltzer}},
  \bibinfo{author}{\bibfnamefont{L.}~\bibnamefont{Childress}},
  \bibnamefont{and} \bibinfo{author}{\bibfnamefont{A.}~\bibnamefont{Gali}},
  \bibinfo{journal}{New Journal of Physics} \textbf{\bibinfo{volume}{13}},
  \bibinfo{pages}{025021} (\bibinfo{year}{2011}).

\bibitem[{\citenamefont{Dr\'eau et~al.}(2012)\citenamefont{Dr\'eau, Maze,
  Lesik, Roch, and Jacques}}]{Dreau2012}
\bibinfo{author}{\bibfnamefont{A.}~\bibnamefont{Dr\'eau}},
  \bibinfo{author}{\bibfnamefont{J.-R.} \bibnamefont{Maze}},
  \bibinfo{author}{\bibfnamefont{M.}~\bibnamefont{Lesik}},
  \bibinfo{author}{\bibfnamefont{J.-F.} \bibnamefont{Roch}}, \bibnamefont{and}
  \bibinfo{author}{\bibfnamefont{V.}~\bibnamefont{Jacques}},
  \bibinfo{journal}{Phys. Rev. B} \textbf{\bibinfo{volume}{85}},
  \bibinfo{pages}{134107} (\bibinfo{year}{2012}).

\bibitem[{\citenamefont{Jelezko et~al.}(2004)\citenamefont{Jelezko, Gaebel,
  Popa, Domhan, Gruber, and Wrachtrup}}]{JEL2004NUC}
\bibinfo{author}{\bibfnamefont{F.}~\bibnamefont{Jelezko}},
  \bibinfo{author}{\bibfnamefont{T.}~\bibnamefont{Gaebel}},
  \bibinfo{author}{\bibfnamefont{I.}~\bibnamefont{Popa}},
  \bibinfo{author}{\bibfnamefont{M.}~\bibnamefont{Domhan}},
  \bibinfo{author}{\bibfnamefont{A.}~\bibnamefont{Gruber}}, \bibnamefont{and}
  \bibinfo{author}{\bibfnamefont{J.}~\bibnamefont{Wrachtrup}},
  \bibinfo{journal}{Phys. Rev. Lett.} \textbf{\bibinfo{volume}{93}},
  \bibinfo{pages}{130501} (\bibinfo{year}{2004}).

\bibitem[{\citenamefont{Neumann et~al.}(2008)\citenamefont{Neumann, Mizuochi,
  Rempp, Hemmer, Watanabe, Yamasaki, Jacques, Gaebel, Jelezko, and
  Wrachtrup}}]{Neumann2008}
\bibinfo{author}{\bibfnamefont{P.}~\bibnamefont{Neumann}},
  \bibinfo{author}{\bibfnamefont{N.}~\bibnamefont{Mizuochi}},
  \bibinfo{author}{\bibfnamefont{F.}~\bibnamefont{Rempp}},
  \bibinfo{author}{\bibfnamefont{P.}~\bibnamefont{Hemmer}},
  \bibinfo{author}{\bibfnamefont{H.}~\bibnamefont{Watanabe}},
  \bibinfo{author}{\bibfnamefont{S.}~\bibnamefont{Yamasaki}},
  \bibinfo{author}{\bibfnamefont{V.}~\bibnamefont{Jacques}},
  \bibinfo{author}{\bibfnamefont{T.}~\bibnamefont{Gaebel}},
  \bibinfo{author}{\bibfnamefont{F.}~\bibnamefont{Jelezko}}, \bibnamefont{and}
  \bibinfo{author}{\bibfnamefont{J.}~\bibnamefont{Wrachtrup}},
  \bibinfo{journal}{Science} \textbf{\bibinfo{volume}{320}},
  \bibinfo{pages}{1326} (\bibinfo{year}{2008}).

\bibitem[{\citenamefont{Maurer et~al.}(2012)\citenamefont{Maurer, Kucsko,
  Latta, Jiang, Yao, Bennett, Pastawski, Hunger, Chisholm, Markham
  et~al.}}]{Maurer2012}
\bibinfo{author}{\bibfnamefont{P.~C.} \bibnamefont{Maurer}},
  \bibinfo{author}{\bibfnamefont{G.}~\bibnamefont{Kucsko}},
  \bibinfo{author}{\bibfnamefont{C.}~\bibnamefont{Latta}},
  \bibinfo{author}{\bibfnamefont{L.}~\bibnamefont{Jiang}},
  \bibinfo{author}{\bibfnamefont{N.~Y.} \bibnamefont{Yao}},
  \bibinfo{author}{\bibfnamefont{S.~D.} \bibnamefont{Bennett}},
  \bibinfo{author}{\bibfnamefont{F.}~\bibnamefont{Pastawski}},
  \bibinfo{author}{\bibfnamefont{D.}~\bibnamefont{Hunger}},
  \bibinfo{author}{\bibfnamefont{N.}~\bibnamefont{Chisholm}},
  \bibinfo{author}{\bibfnamefont{M.}~\bibnamefont{Markham}},
  \bibnamefont{et~al.}, \bibinfo{journal}{Science}
  \textbf{\bibinfo{volume}{336}}, \bibinfo{pages}{1283} (\bibinfo{year}{2012}).

\bibitem[{\citenamefont{van Oort et~al.}(1990)\citenamefont{van Oort, Stroomer,
  and Glasbeek}}]{vanOort90}
\bibinfo{author}{\bibfnamefont{E.}~\bibnamefont{van Oort}},
  \bibinfo{author}{\bibfnamefont{P.}~\bibnamefont{Stroomer}}, \bibnamefont{and}
  \bibinfo{author}{\bibfnamefont{M.}~\bibnamefont{Glasbeek}},
  \bibinfo{journal}{Phys. Rev. B} \textbf{\bibinfo{volume}{42}},
  \bibinfo{pages}{8605} (\bibinfo{year}{1990}).

\bibitem[{\citenamefont{van Oort et~al.}(1990)\citenamefont{van Oort}}]{trial}
\bibinfo{author}{\bibfnamefont{E.}~\bibnamefont{van Oort}},
  \bibinfo{journal}{Ph.D. thesis, University of Amsterdam,}\bibinfo{year}{ 1990}.

\bibitem[{\citenamefont{Jensen et~al.}(2013)\citenamefont{Jensen, Acosta,
  Jarmola, and Budker}}]{JEN2012}
\bibinfo{author}{\bibfnamefont{K.}~\bibnamefont{Jensen}},
  \bibinfo{author}{\bibfnamefont{V.~M.} \bibnamefont{Acosta}},
  \bibinfo{author}{\bibfnamefont{A.}~\bibnamefont{Jarmola}}, \bibnamefont{and}
  \bibinfo{author}{\bibfnamefont{D.}~\bibnamefont{Budker}},
  \bibinfo{journal}{Phys. Rev. B} \textbf{\bibinfo{volume}{87}},
  \bibinfo{pages}{014115} (\bibinfo{year}{2013}).

\bibitem[{\citenamefont{Baranov et~al.}(2011)\citenamefont{Baranov, Soltamova,
  Tolmachev, Romanov, Babunts, Shakhov, Kidalov, Vul’, Mamin, Orlinskii
  et~al.}}]{BAR2011}
\bibinfo{author}{\bibfnamefont{P.~G.} \bibnamefont{Baranov}},
  \bibinfo{author}{\bibfnamefont{A.~A.} \bibnamefont{Soltamova}},
  \bibinfo{author}{\bibfnamefont{D.~O.} \bibnamefont{Tolmachev}},
  \bibinfo{author}{\bibfnamefont{N.~G.} \bibnamefont{Romanov}},
  \bibinfo{author}{\bibfnamefont{R.~A.} \bibnamefont{Babunts}},
  \bibinfo{author}{\bibfnamefont{F.~M.} \bibnamefont{Shakhov}},
  \bibinfo{author}{\bibfnamefont{S.~V.} \bibnamefont{Kidalov}},
  \bibinfo{author}{\bibfnamefont{A.~Y.} \bibnamefont{Vul’}},
  \bibinfo{author}{\bibfnamefont{G.~V.} \bibnamefont{Mamin}},
  \bibinfo{author}{\bibfnamefont{S.~B.} \bibnamefont{Orlinskii}},
  \bibnamefont{et~al.}, \bibinfo{journal}{Small} \textbf{\bibinfo{volume}{7}},
  \bibinfo{pages}{1533} (\bibinfo{year}{2011}).

\bibitem[{\citenamefont{Babunts et~al.}(2012)\citenamefont{Babunts, Soltamova,
  Tolmachev, Soltamov, Gurin, Anisimov, Preobrazhenskii, and
  Baranovi}}]{Babunts2012}
\bibinfo{author}{\bibfnamefont{R.}~\bibnamefont{Babunts}},
  \bibinfo{author}{\bibfnamefont{A.}~\bibnamefont{Soltamova}},
  \bibinfo{author}{\bibfnamefont{D.}~\bibnamefont{Tolmachev}},
  \bibinfo{author}{\bibfnamefont{V.}~\bibnamefont{Soltamov}},
  \bibinfo{author}{\bibfnamefont{A.}~\bibnamefont{Gurin}},
  \bibinfo{author}{\bibfnamefont{A.}~\bibnamefont{Anisimov}},
  \bibinfo{author}{\bibfnamefont{V.}~\bibnamefont{Preobrazhenskii}},
  \bibnamefont{and} \bibinfo{author}{\bibfnamefont{P.}~\bibnamefont{Baranovi}},
  \bibinfo{journal}{JETP Letters} \textbf{\bibinfo{volume}{95}},
  \bibinfo{pages}{429} (\bibinfo{year}{2012}).

\bibitem[{\citenamefont{Acosta et~al.}(2010)\citenamefont{Acosta, Jarmola,
  Bauch, and Budker}}]{ACO2010PRB}
\bibinfo{author}{\bibfnamefont{V.~M.} \bibnamefont{Acosta}},
  \bibinfo{author}{\bibfnamefont{A.}~\bibnamefont{Jarmola}},
  \bibinfo{author}{\bibfnamefont{E.}~\bibnamefont{Bauch}}, \bibnamefont{and}
  \bibinfo{author}{\bibfnamefont{D.}~\bibnamefont{Budker}},
  \bibinfo{journal}{Physical Review B} \textbf{\bibinfo{volume}{82}},
  \bibinfo{pages}{201202} (\bibinfo{year}{2010}).

\bibitem[{\citenamefont{Balmer et~al.}(2009)\citenamefont{Balmer, Brandon,
  Clewes, Dhillon, Dodson, Friel, Inglis, Madgwick, Markham, Mollart
  et~al.}}]{BAL2009CVD}
\bibinfo{author}{\bibfnamefont{R.~S.} \bibnamefont{Balmer}},
  \bibinfo{author}{\bibfnamefont{J.~R.} \bibnamefont{Brandon}},
  \bibinfo{author}{\bibfnamefont{S.~L.} \bibnamefont{Clewes}},
  \bibinfo{author}{\bibfnamefont{H.~K.} \bibnamefont{Dhillon}},
  \bibinfo{author}{\bibfnamefont{J.~M.} \bibnamefont{Dodson}},
  \bibinfo{author}{\bibfnamefont{I.}~\bibnamefont{Friel}},
  \bibinfo{author}{\bibfnamefont{P.~N.} \bibnamefont{Inglis}},
  \bibinfo{author}{\bibfnamefont{T.~D.} \bibnamefont{Madgwick}},
  \bibinfo{author}{\bibfnamefont{M.~L.} \bibnamefont{Markham}},
  \bibinfo{author}{\bibfnamefont{T.~P.} \bibnamefont{Mollart}},
  \bibnamefont{et~al.}, \bibinfo{journal}{Journal of Physics: Condensed Matter}
  \textbf{\bibinfo{volume}{21}}, \bibinfo{pages}{364221}
  (\bibinfo{year}{2009}).

\bibitem[{\citenamefont{Takahashi et~al.}(2008)\citenamefont{Takahashi, Hanson,
  van Tol, Sherwin, and Awschalom}}]{TAK2008}
\bibinfo{author}{\bibfnamefont{S.}~\bibnamefont{Takahashi}},
  \bibinfo{author}{\bibfnamefont{R.}~\bibnamefont{Hanson}},
  \bibinfo{author}{\bibfnamefont{J.}~\bibnamefont{van Tol}},
  \bibinfo{author}{\bibfnamefont{M.~S.} \bibnamefont{Sherwin}},
  \bibnamefont{and} \bibinfo{author}{\bibfnamefont{D.~D.}
  \bibnamefont{Awschalom}}, \bibinfo{journal}{Physical Review Letters}
  \textbf{\bibinfo{volume}{101}}, \bibinfo{pages}{047601}
  (\bibinfo{year}{2008}).
  
  \bibitem[{\citenamefont{Belthangady et~al.}(2013)\citenamefont{Belthangady, Bar-Gill, Pham, Arai, Le Sage, Cappellaro and Walsworth}}]{BEL2013}
\bibinfo{author}{\bibfnamefont{C.}~\bibnamefont{Belthangady}},
  \bibinfo{author}{\bibfnamefont{N.}~\bibnamefont{Bar-Gill}},
  \bibinfo{author}{\bibfnamefont{L.~M.}~\bibnamefont{Pham}},
  \bibinfo{author}{\bibfnamefont{K.} \bibnamefont{Arai}},
    \bibinfo{author}{\bibfnamefont{D.}~\bibnamefont{Le Sage}},
  \bibinfo{author}{\bibfnamefont{P.} \bibnamefont{Cappellaro}},
  \bibnamefont{and} \bibinfo{author}{\bibfnamefont{R.~L.}
  \bibnamefont{Walsworth}}, \bibinfo{journal}{Phys. Rev. Lett.}
  \textbf{\bibinfo{volume}{110}}, \bibinfo{pages}{157601}
  (\bibinfo{year}{2013}).

\bibitem[{\citenamefont{Jarmola et~al.}(2012)\citenamefont{Jarmola, Acosta,
  Jensen, Chemerisov, and Budker}}]{JAR2012}
\bibinfo{author}{\bibfnamefont{A.}~\bibnamefont{Jarmola}},
  \bibinfo{author}{\bibfnamefont{V.~M.} \bibnamefont{Acosta}},
  \bibinfo{author}{\bibfnamefont{K.}~\bibnamefont{Jensen}},
  \bibinfo{author}{\bibfnamefont{S.}~\bibnamefont{Chemerisov}},
  \bibnamefont{and} \bibinfo{author}{\bibfnamefont{D.}~\bibnamefont{Budker}},
  \bibinfo{journal}{Phys. Rev. Lett.} \textbf{\bibinfo{volume}{108}},
  \bibinfo{pages}{197601} (\bibinfo{year}{2012}).

\bibitem[{\citenamefont{Pham et~al.}(2012)\citenamefont{Pham, Bar-Gill,
  Belthangady, Le~Sage, Cappellaro, Lukin, Yacoby, and Walsworth}}]{PHA2012}
\bibinfo{author}{\bibfnamefont{L.~M.} \bibnamefont{Pham}},
  \bibinfo{author}{\bibfnamefont{N.}~\bibnamefont{Bar-Gill}},
  \bibinfo{author}{\bibfnamefont{C.}~\bibnamefont{Belthangady}},
  \bibinfo{author}{\bibfnamefont{D.}~\bibnamefont{Le~Sage}},
  \bibinfo{author}{\bibfnamefont{P.}~\bibnamefont{Cappellaro}},
  \bibinfo{author}{\bibfnamefont{M.~D.} \bibnamefont{Lukin}},
  \bibinfo{author}{\bibfnamefont{A.}~\bibnamefont{Yacoby}}, \bibnamefont{and}
  \bibinfo{author}{\bibfnamefont{R.~L.} \bibnamefont{Walsworth}},
  \bibinfo{journal}{Phys. Rev. B} \textbf{\bibinfo{volume}{86}},
  \bibinfo{pages}{045214} (\bibinfo{year}{2012}).

\bibitem[{\citenamefont{Stanwix et~al.}(2010)\citenamefont{Stanwix, Pham, Maze,
  Le~Sage, Yeung, Cappellaro, Hemmer, Yacoby, Lukin, and Walsworth}}]{STA2010}
\bibinfo{author}{\bibfnamefont{P.~L.} \bibnamefont{Stanwix}},
  \bibinfo{author}{\bibfnamefont{L.~M.} \bibnamefont{Pham}},
  \bibinfo{author}{\bibfnamefont{J.~R.} \bibnamefont{Maze}},
  \bibinfo{author}{\bibfnamefont{D.}~\bibnamefont{Le~Sage}},
  \bibinfo{author}{\bibfnamefont{T.~K.} \bibnamefont{Yeung}},
  \bibinfo{author}{\bibfnamefont{P.}~\bibnamefont{Cappellaro}},
  \bibinfo{author}{\bibfnamefont{P.~R.} \bibnamefont{Hemmer}},
  \bibinfo{author}{\bibfnamefont{A.}~\bibnamefont{Yacoby}},
  \bibinfo{author}{\bibfnamefont{M.~D.} \bibnamefont{Lukin}}, \bibnamefont{and}
  \bibinfo{author}{\bibfnamefont{R.~L.} \bibnamefont{Walsworth}},
  \bibinfo{journal}{Physical Review B} \textbf{\bibinfo{volume}{82}},
  \bibinfo{pages}{201201} (\bibinfo{year}{2010}).

\bibitem[{\citenamefont{Smith et~al.}(1959)\citenamefont{Smith, Sorokin,
  Gelles, and Lasher}}]{SMI1959}
\bibinfo{author}{\bibfnamefont{W.~V.} \bibnamefont{Smith}},
  \bibinfo{author}{\bibfnamefont{P.~P.} \bibnamefont{Sorokin}},
  \bibinfo{author}{\bibfnamefont{I.~L.} \bibnamefont{Gelles}},
  \bibnamefont{and} \bibinfo{author}{\bibfnamefont{G.~J.}
  \bibnamefont{Lasher}}, \bibinfo{journal}{Physical Review}
  \textbf{\bibinfo{volume}{115}}, \bibinfo{pages}{1546} (\bibinfo{year}{1959}).

\bibitem[{\citenamefont{Gali et~al.}(2008)\citenamefont{Gali, Fyta, and Kaxiras}}]{GAL2008}
\bibinfo{author}{\bibfnamefont{A.} \bibnamefont{Gali}},
  \bibinfo{author}{\bibfnamefont{M.} \bibnamefont{Fyta}},
  \bibnamefont{and} \bibinfo{author}{\bibfnamefont{E.}
  \bibnamefont{Kaxiras}}, \bibinfo{journal}{Phys. Rev. B}
  \textbf{\bibinfo{volume}{77}}, \bibinfo{pages}{155206} (\bibinfo{year}{2008}).

\end{thebibliography}

\end{document}